\documentclass[useAMS,usenatbib]{mn2e}
\usepackage{graphicx}
\usepackage{verbatim}
\usepackage{times}

\newcommand{\ie}{{i.e.}}

\newcommand{\etalr}{{et~al.}}

\newcommand{\beq}{\begin{equation}}
\newcommand{\eeq}{\end{equation}}
\newcommand{\bea}{\begin{eqnarray}}
\newcommand{\eea}{\end{eqnarray}}
\newcommand{\lsim}{\,\lower2truept\hbox{${<\atop\hbox{\raise4truept\hbox{$\sim$}}}$}\,}
\newcommand{\gsim}{\,\lower2truept\hbox{${>\atop\hbox{\raise4truept\hbox{$\sim$}}}$}\,}
\newcommand{\wjjj}[6]
{{
\left(
\begin{array}{lcr} #1 & #2 & #3 \\#4 & #5 & #6 \end{array}
\right)
}}

\title[CMB Power Spectrum Estimation with Destriping]
{Cosmic microwave background power spectrum estimation with the
destriping technique}

 \author[T. Poutanen, D. Maino, H. Kurki-Suonio, E. Keih\"anen, E. Hivon]
 {T. Poutanen$^{1}$\thanks{E-mail: torsti.poutanen@helsinki.fi}, D. Maino$^{2}$, H.
Kurki-Suonio$^{3}$,
  E. Keih\"anen$^{1,3}$, E. Hivon$^{4}$\\
 $^1$ Helsinki Institute of Physics, P.O. Box 64, FIN-00014, Helsinki, Finland \\
 $^2$ Dipartimento di Fisica, Universit\`{a} di Milano, Via Celoria 16,
  I-20131 Milano, Italy\\
 $^3$ University of Helsinki, Department of Physical Sciences, P.O. Box 64,
  FIN-00014, Helsinki, Finland\\
 $^4$ IPAC, MS 100-22, Caltech, Pasadena, CA 91125}

\pagerange{\pageref{firstpage}--\pageref{lastpage}} \pubyear{2004}

\begin{document}

\maketitle

\label{firstpage}
\begin{abstract}
Extraction of the CMB (Cosmic Microwave Background) angular power spectrum is a
challenging task for current and future CMB experiments due to the large data
sets involved. Here we describe
an implementation of MASTER (Monte carlo Apodised Spherical
Transform EstimatoR) described in \citet{HIV02} which exploits the
destriping technique
as a map-making method.
In this method a noise estimate based on destriped noise-only MC (Monte Carlo)
simulations is subtracted from the pseudo angular power spectrum. As a working
case we use realistic simulations of the {\it PLANCK} LFI (Low Frequency
Instrument). We found that the effect of destriping on a pure sky signal is
minimal and requires no correction.  Instead we found an effect related to the
distribution of detector pointings,
which affects the
high-$\ell$ part of the power spectrum. We correct for this by subtracting a
``signal bias'' estimated by MC simulations.  We also give analytical estimates
for this signal bias.  Our method is fast and accurate enough (the estimator is
un-biased and errors are close to theoretical expectations for maximal
accuracy) to estimate the CMB angular power spectra for current and future CMB
space missions.  This study is related to {\it PLANCK} LFI activities.
\end{abstract}

\begin{keywords}
methods: data analysis -- cosmology: cosmic microwave background.
\end{keywords}

\section{Introduction}
\label{sec:intro}

In the favoured model of structure formation driven by inflation,
primordial fluctuations are expected to be Gaussian distributed.
In this case all the statistical information encoded into CMB
(Cosmic Microwave Background) anisotropies is completely described
by their angular power spectrum $C_\ell$. The main issue in this
paper is the extraction of the $C_\ell$ values starting from an
observed CMB map. In recent years a maximum likelihood approach to
the problem has been developed \citep{GOR96} and successfully
applied to the {\it COBE} - DMR data \citep{BEN96}. This process,
however, involves a number of operations scaling as $\sim N_{\rm
pix}^3$, where $N_{\rm pix}$ is the number of pixels that cover
the sky.

Recent ground-based and balloon-borne experiments have improved
our knowledge of the CMB anisotropy due to the better angular
resolution and higher sensitivity of these experiments. Due to the
sizes of their data sets they have already posed challenges in
extracting an unbiased estimate of $C_\ell$. This will become more
demanding with current and future satellite experiments like {\it
WMAP} \citep{BEN03} and {\it PLANCK} \citep{TAU00} which produce
maps with $N_{\rm pix} \simeq {\rm few}\times 10^6$. Therefore,
brute force maximum likelihood estimation of the CMB angular power
spectrum is not feasible without approximations or assumptions
specific to particular experimental or observing strategies
\citep{OHS99,WAN03}.

New methods are under study and some of them have already produced
interesting results. \citet{HIV02} introduced the MASTER technique
used to construct the BOOMERanG angular power spectrum in
\citet{NET02}. An extension of MASTER is being prepared to treat
{\it PLANCK}-like data sets. The core of the MASTER technique is
{\it i)} a high-pass filter applied to the TOD (Time Ordered Data)
as part of a naive map-making process in order to reduce
instrumental non-white noise, and {\it ii)} MC (Monte Carlo)
simulations of pure signal, pure noise, and signal plus noise to
correct for data filtering and instrumental noise, and for
estimating the error bars of $C_\ell$. The approach has been shown
to be unbiased and it gives error bars close to theoretical ones
when cosmic variance and instrumental noise properties are
concerned \citep{HIV02}.

\citet{BAL02} follow a MASTER-like approach but instead of using a
high-pass filter they consider an IGLS (Iterative Generalized
Least Square) map-making algorithm already developed to treat {\it
PLANCK} data \citep{NAT01}. Furthermore, they estimate noise
properties needed for proper MC simulations directly from the data
\citep{NAT02}. The extracted $C_\ell$ are unbiased and have close
to optimal error bars.

In this paper we present an approach similar to that of
\citet{HIV02} but exploiting the destriping algorithm
\citep{BUR97,DEL98,MAI99a,MAI99b} in the map-making process.
Although destriping algorithms are considered approximations of
proper IGLS map-making in the sense that they will not necessarily
produce the minimum variance map, they are able to provide
estimates of various systematic effects, remove drifts from TOD
and return cleaned TOD (e.g., see \citealt{MEN02} for an
application with periodic fluctuations). Furthermore, destriping
makes no assumptions on the beam shape. It has been demonstrated
\citep{MAI02} that the instrument noise is the driver of the
destriping performance regardless of the beam shape (including
sidelobes).

A generalized maximum likelihood approach to the destriping method
has been implemented \citep{KEI03} which is able to fit different
sets of base functions (in addition to the simple constant
baseline) and, in principle, could better remove the contributions
of different systematic effects from the TODs. All these
properties make destriping attractive. Additionally, it is fast
and needs no prior information on the instrumental noise. Note,
however, that as we combine destriping with the MASTER approach
for the $C_\ell$ estimation, this part utilizes information on
noise properties. Both iterative and non-iterative methods to
estimate the noise characteristics directly from the data have
been proposed \citep{DOR01,NAT02}. These methods have been applied
in the $C_\ell$ estimation \citep[e.g.][]{NET02,BAL02}.

This paper is organised as follows. In Section~\ref{sec:destri} we
describe the destriping technique. In Section~\ref{sec:MASTER} we
review the MASTER approach for the extraction of $C_\ell$ from a
map. We have then applied our combination of destriping with the
MASTER approach to simulations of one {\it PLANCK} LFI 100 GHz
detector. The applied scanning strategy and sky coverages are
described in Section~\ref{sec:scanning_strategy}. In
Section~\ref{sec:filter_function} we explain our findings
concerning the filter function and signal bias that are possible
means to model the effects caused by our map-making method on the
spectrum estimates. The analysis pipeline and the CPU times
required to run it are explained in Section~\ref{sec:times}.
Section~\ref{sec:simul} presents the simulations and the
simulation results. We draw our conclusions in
Section~\ref{sec:conclusion}. In
Appendix~\ref{sec:pointing_distribution} we describe in more
details the signal bias approach for modelling the effect of the
distribution of detector pointings in output map pixels.

\section{Destriping Technique} \label{sec:destri}

The destriping technique for map making has been derived from the
COBRAS/SAMBA Phase-A study \citep{BER96} and was originally
implemented by \citet{BUR97} and by \citet{MAI99b}. (See also
\citealt{DEL98}.) Developed for the {\it PLANCK} satellite it
makes use of the fact that {\it PLANCK} is a spinning spacecraft.
Detector beams are drawing almost great circles on the
sky. Each scan circle is observed 60 times before the spin axis is repointed.
In order to reduce the level of instrumental noise, the signal can be averaged
over these 60 scan circles. Although the destriping technique has been
developed in this framework, it could in principle be applied to any experiment
with an observing strategy with repeated measurements of the same regions of
the sky.

As pointed out by \citet{JAN96}, the effect of the instrumental
noise, in particular $1/f$ noise, on the average scan circle
signal can be approximated by a uniform offset or ``baseline''.
The key problem in destriping is to find the magnitudes of the
baselines. The destriping technique uses the redundancy of the
observing strategy, \ie\ it considers all intersections (crossing
points) between the scan circles, to obtain these
magnitudes. 
We identify a crossing point when two
points from different scan circles fall inside the same pixel in
the sky. 
The small scale structure in the signal will cause destriping to
insert artefacts at some level when the pointings of the observing
beam do not fall exactly on each other within the pixels of the
crossing point search. We will come back to this point later in
Section~\ref{sec:filter_function}. In this study we utilize the
HEALPix\footnote{http://www.eso.org/science/healpix} pixelisation
scheme \citep{GOR99}. Its pixel dimension is set by the $N_{\rm
side}$ resolution parameter.  A map of the full sky contains
$12N_{\rm side}^2$ pixels.

After the baseline magnitudes have been recovered they can be used to subtract
an estimate of the instrumental $1/f$ noise from the observed TOD. The final
sky map is built from this cleaned TOD by simply coadding (averaging) the
observations falling in each sky
pixel. 
Some residual $1/f$ noise will remain
after the subtraction of the baselines. 
However, for knee frequencies $f_k \lsim 0.4$~Hz the uniform
baseline approximation works well and the level of residual noise
stripes in the output map is well below the white noise level
\citep{MAI02}. The knee frequency considered in this study ($f_k =
0.1$~Hz) lies well below the above limit.

\section{The MASTER Approach}
\label{sec:MASTER}

The CMB temperature field observed over the whole sky can be expanded in
spherical harmonics $Y_{\ell m}$,
 \beq
   \Delta T(\mathbf{n}) =
   \sum_{\ell m} a_{\ell m} Y_{\ell m}(\mathbf{n})\, ,
 \label{harmofs}
 \eeq
where the $Y_{\ell m}$ form a complete orthogonal set of functions over the
sky. Here $\mathbf{n}$ is a unit vector pointing to a location on the sky. If
CMB temperature fluctuations are assumed to be Gaussian distributed, the
expansion coefficients $a_{\ell m}$ are Gaussian distributed complex random
variables with expectation value
 \beq
   \langle a_{\ell m} \rangle = 0\, ,
 \eeq
and correlation
 \beq
   \langle a_{\ell m} a_{\ell' m'}^\star \rangle = \delta_{\ell \ell'}
   \delta_{m m'} C_\ell^{\rm th}\, \label{clth},
 \eeq
where $C_\ell^{\rm th}$ is
the underlying ``theoretical'' angular power spectrum of the CMB.

In the case of full sky observations an unbiased estimator of
$C_\ell^{\rm th}$ is given by \beq
   C_\ell = \frac{1}{2\ell +1}
   \sum_{m=-\ell}^{\ell} |a_{\ell m}|^2\, . \label{clunb}
\eeq The coefficients $C_\ell$ are $\chi_{\nu}^2$ distributed with
mean equal to the theoretical value $C_\ell^{\rm th}$ and standard
deviation, ``cosmic variance'', given by $\Delta C_\ell =
\sqrt{2/(2\ell +1)}C_\ell^{\rm th}$ \citep[e.g.][]{KNO95}. Here
$\nu = 2\ell +1$ is the number of degrees of freedom corresponding
to a given multipole number $\ell$.

From a pixelised map of the full sky the coefficients
$\widetilde{a}_{\ell m}$ are estimated as a sum over pixels
 \beq
    \widetilde{a}_{\ell m} = \Omega_p\sum_pw_pT_pY_{\ell m}^\star\left(\mathbf{n}_p\right) \,,
 \label{alm}
 \eeq
where $\Omega_p$ is the pixel area (same for all pixels in
HEALPix), $T_p$ the pixel temperature, $\mathbf{n}_p$ the unit
vector pointing to the centre of the pixel, and $w_p$ the pixel
weight. Throughout this study the pixel weights equal one for all
observed pixels.

Including the effect of instrumental noise and a symmetric antenna beam, the
expectation value of the ``observed'' angular power spectrum
$\widetilde{C}_\ell$ is (by inserting $\widetilde{a}_{\ell m}$ to Eq.
(\ref{clunb}))
 \beq
   \langle\widetilde{C}_\ell\rangle = B^2_\ell C_\ell^{\rm th} +
   \langle\widetilde{N}_\ell\rangle\, ,
 \eeq
and variance
 \beq
   \Delta^2\widetilde{C}_\ell
    = \frac{2}{2\ell+1}\left(B_\ell^2C_\ell^{\rm th}
   + \langle\widetilde{N}_\ell\rangle\right)^2\, ,
 \eeq
where $B_\ell$ is the shape of the beam in $\ell$ space, and
$\langle\widetilde{N}_\ell\rangle$ is the expectation value of the
angular power spectrum of the noise projected onto the sky.

Thus an unbiased estimate of $C_\ell^{\rm th}$ is given by
 \beq
   \widehat{C}_\ell = B_\ell^{-2}\left(\widetilde{C}_\ell
   - \langle\widetilde{N}_\ell\rangle\right),
 \eeq
which has the expectation value $\langle\widehat{C}_\ell\rangle = C_\ell^{\rm th}$
and variance
 \beq
   \Delta^2\widehat{C}_\ell = \frac{2}{2\ell+1}\left(C_\ell^{\rm th}
   + \frac{\langle\widetilde{N}_\ell\rangle}{B_\ell^2}\right)^2\, .
 \label{thvariance}
 \eeq

The situation in a real experiment is more complicated since the
entire sky may not be available for the observation. This may
happen due to partial sky coverage of the scanning, removal of
corrupted data or due to cutting away the galactic plane region.
The pixels are therefore weighted by a window function
$W(\mathbf{n})$, that states which pixels are observed and which
are not, in a top-hat fashion (used here) or using a more complex
apodization scheme. On the cut sky the $Y_{\ell m}$ are no more an
orthogonal set of functions and a correlation between the $\ell$
modes is introduced. However, it is still possible to determine
the expansion coefficients $\widetilde{a}_{\ell m}$ over the cut
sky, by summing over the observed pixels in Eq.\ (\ref{alm}). The
pseudo power spectrum $\widetilde{C}_\ell$ can be obtained by
inserting the coefficients to Eq. (\ref{clunb}). The expectation
value of the pseudo power spectrum can be related to the true
angular power spectrum $C_\ell^{th}$ of the sky \citep{HIV02}
 \beq
    \langle \widetilde{C}_\ell\rangle =
    \sum_{\ell'} M_{\ell \ell'} C_{\ell'}^{th},
 \label{newclunb}
 \eeq
where the kernel matrix $M_{\ell \ell'}$ describes the mode-mode coupling due
to the cut sky. For the top-hat window function this matrix depends only on the
geometry of the cut applied to the data.

In our method the temperature of a pixel of the output map is an average of all
cleaned (baselines subtracted) observations falling in that pixel. The
pointings of the observations are distributed across the pixel area.
The distribution of the observations is likely to cause some
distortion in the observed CMB angular power spectrum, that may
need to be corrected in the power spectrum estimation stage. In
this study the signal TOD is generated by scanning a pixelized sky
map realization. The spread of the pointings is modelled by
picking the temperatures from a map (the {\em sky map}) with a
higher resolution than the resolution of the final {\em output
map}.  For this study the pointings of the scan circles between
successive satellite spin axis repointings are assumed to fall
exactly on top of each other. In real experiment this will not
happen due to various non-idealities in the satellite motion (e.g.
nutation of the spin axis pointing, \citealt{LEE02}). The way we
model the pointing distribution may exaggerate the distortion of
the estimated CMB angular power spectrum.

In the original MASTER implementation \citep{HIV02} a high-pass
filter was applied to the TOD to reduce low frequency noise.  The
effect of this filtering on the pseudo angular power spectrum was
modelled by a filter transfer function $F_\ell$, so that we would
have an approximate relation
 \beq
   \langle \widetilde{C}_\ell\rangle = \sum_{\ell'}M_{\ell
   \ell'} F_{\ell'} B^2_{\ell'} C_{\ell'}^{th} + \langle
   \widetilde{N}_\ell \rangle\, , \label{clgen}
   \label{clgeb}
 \eeq
The filter function $F_\ell$ is to be determined by signal-only MC simulations.

In our approach the low-frequency noise is removed by destriping,
so there is no need for a high-pass filter.  However, our
map-making method may still have some effect on the signal, that
needs to be corrected for.  In Section~\ref{sec:filter_function}
we study whether this effect could also be modelled with a filter
function in the way of Eq.\ (\ref{clgen}). Then the
$\widetilde{C}_\ell$ is the pseudo power spectrum obtained from
the destriped map, $B_\ell^2$ is a function including the beam
response and the effect of finite output map pixel size, and
$\langle \widetilde{N}_\ell\rangle$ is the average of the noise
angular power spectrum (on the cut sky, destriped). In
Section~\ref{sec:filter_function} we introduce an alternative
(``signal bias'') which appears more suitable for our case.

In a real CMB experiment a single noisy realization of the pseudo
power spectrum $\widetilde{C}_\ell$ is available for the
estimation of the true angular power spectrum $C_\ell^{th}$ of the
sky. In this case Eq. (\ref{clgen}) can be inverted to obtain a
formula for the estimate of $C_{\ell}^{th}$,
 \beq
   \widehat{C}_\ell = \frac{\sum_{\ell'}{M_{\ell \ell'}^{-1}
   \left(\widetilde{C}_{\ell'} - \langle \widetilde{N}_{\ell'}
   \rangle \right)}}{F_\ell B^2_\ell}, \label{clestimate}
 \eeq
in the filter function approach. Using the cut sky introduces correlations
between the $\widehat{C}_\ell$.

\subsection{Coupling Kernel}
\label{subsec:kernel}

The mode-mode coupling kernel $M_{\ell \ell'}$ is completely
determined once the window function $W(\mathbf{n})$ is specified.
\citet{HIV02} have demonstrated that the kernel depends only on
the angular power spectrum $W_\ell$ of the sky window function
applied to the data,
 \beq
   M_{\ell_1 \ell_2} =
   \frac{2\ell_2+1}{4\pi}\sum_{\ell_3}(2\ell_3+1)W_{\ell_3}
   \wjjj{\ell_1}{\ell_2}{\ell_3}{0}{0}{0}^{2} , \label{kernel}
 \eeq
where $\wjjj{\ell_1}{\ell_2}{\ell_3}{m_1}{m_2}{m_3}$ is the Wigner 3$-j$
symbol. Therefore, once the window function $W(\mathbf{n})$ applied to the data
is given it is a straightforward task to compute the corresponding coupling
kernel $M_{\ell_1 \ell_2}$. Furthermore, the kernel is the same for all the MC
simulations involved in our technique and needs to be computed only once.

\subsection{Modelling of the Instrument Noise}
\label{subsec:noisePS}

A vital part of the MASTER approach is the estimation of the
angular power spectrum of the instrument noise $\langle
\widetilde{N}_\ell \rangle$ (see Eqs.~(\ref{clgen}) and
(\ref{clestimate})). If the estimate of $\langle
\widetilde{N}_\ell \rangle$ is incorrect a biased $C_\ell$
estimate will be obtained.

Several methods have been developed to estimate noise properties
directly from the data by making use of both iterative
\citep{DOR01} and non-iterative algorithms \citep{NAT02}. If noise
properties can be derived from the data and reasonably modelled,
it is possible to estimate $\langle \widetilde{N}_\ell \rangle$ by
making use of MC realizations. In this approach a set of $N_{\rm
MC}^{\rm(n)}$ pure-noise TOD realizations are generated. They are
then processed with destriping and coadded into maps with galactic
cut applied (if relevant). Thereafter, the angular power spectra
$\widetilde{N}_\ell$ will be extracted. Finally we can construct
the MC average that can be used as an estimate of
$\langle\widetilde{N_\ell}\rangle$ in Eq. (\ref{clestimate}).

\begin{figure}
\begin{center}
\includegraphics[width=8cm,height=4.8cm]{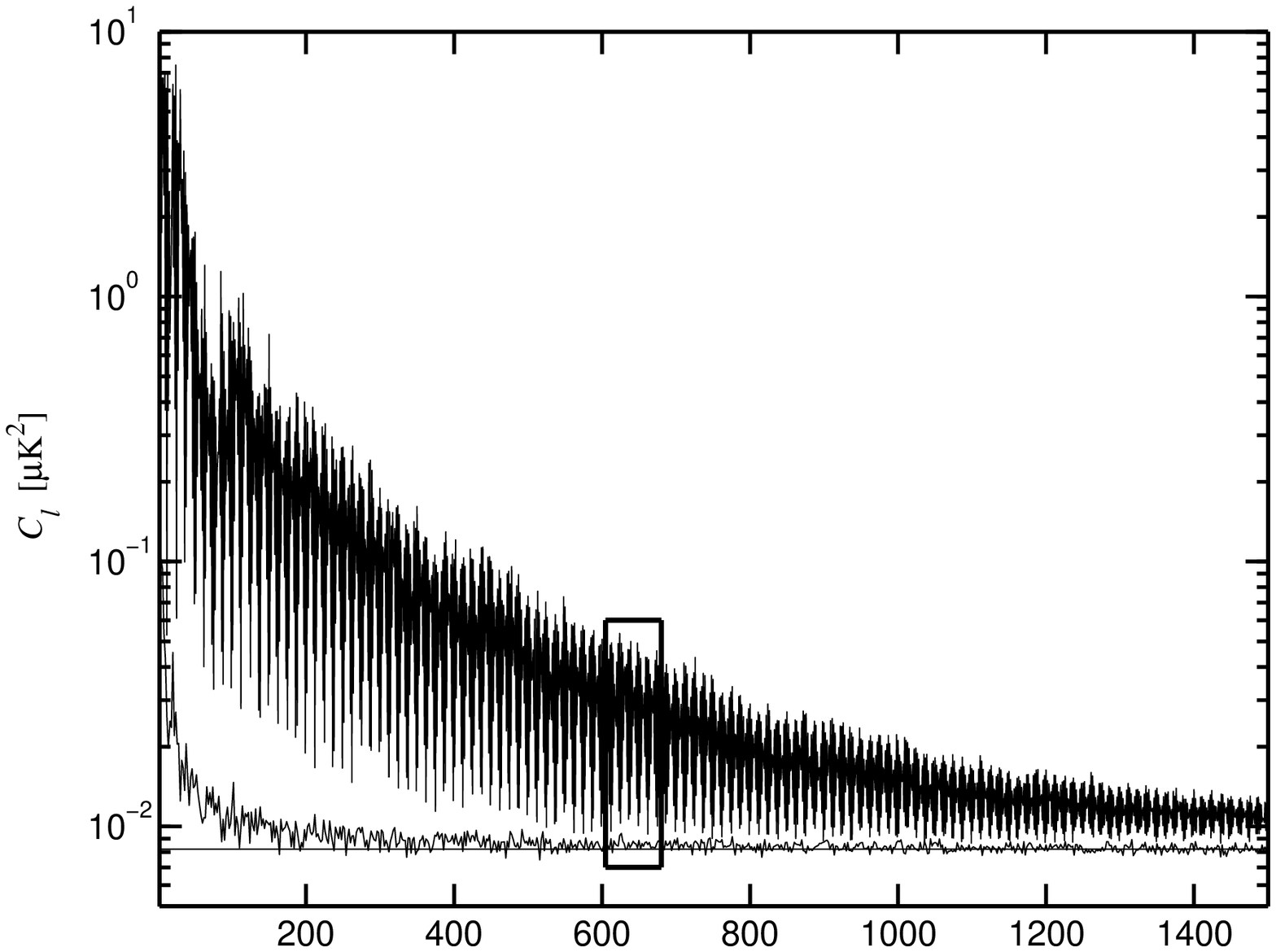}
\includegraphics[width=8cm,height=4.8cm]{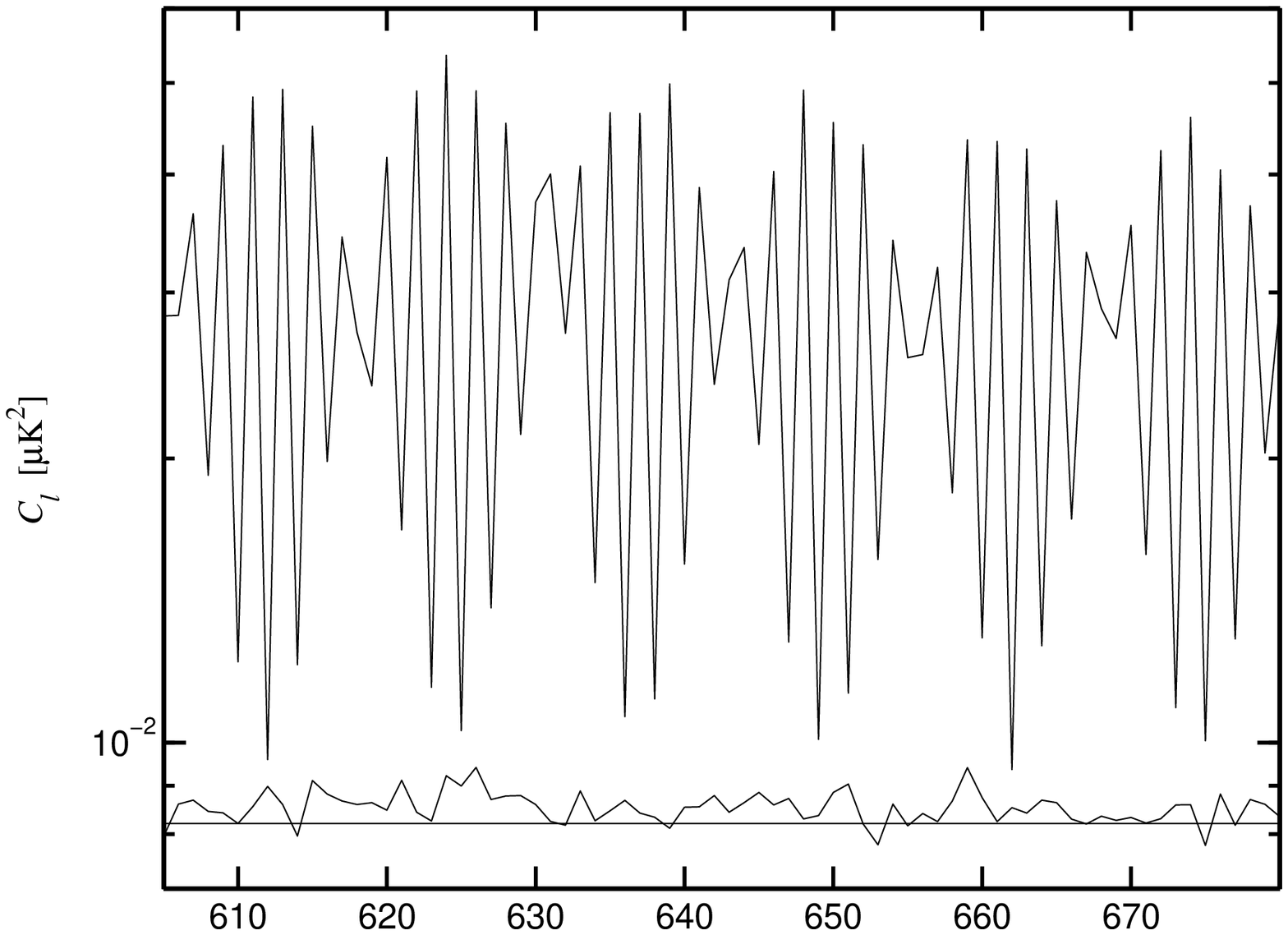}
\includegraphics[width=8cm,height=3.8cm]{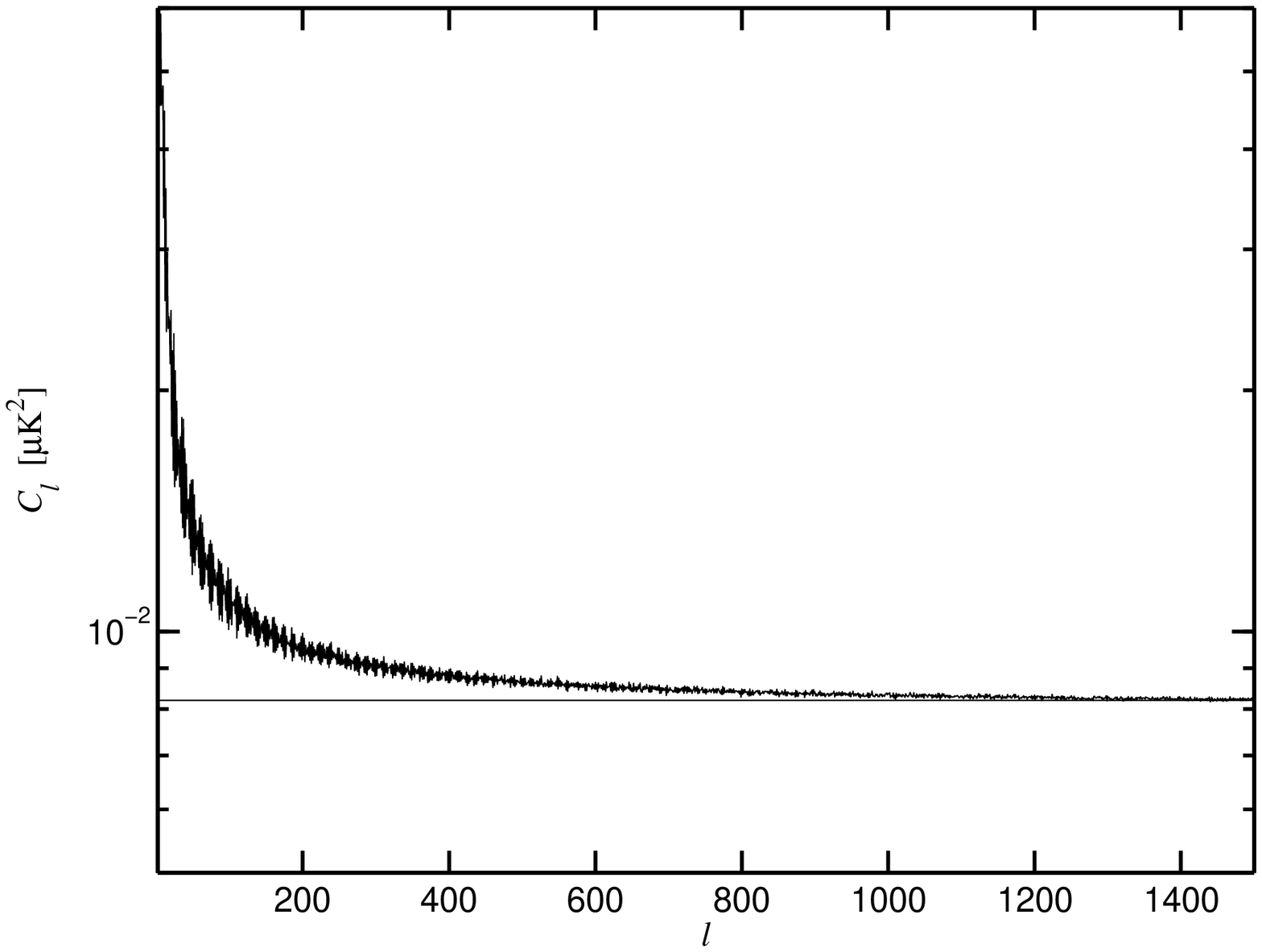}
\end{center}
 \caption{Angular power spectrum of the instrument noise. {\bf Top panel}: Solid
horizontal line is the theoretical white noise spectrum. The upper curve is the
noise spectrum including $1/f$ noise for a single noise realization and no
destriping (naive coaddition from TOD to map). The middle curve is the power
spectrum after destriping. {\bf Middle panel}: The section indicated by the box
(top panel) is shown here. Six blobs are evident in the no-destriping curve.
{\bf Bottom panel}: The top curve is the mean power spectrum (of 100 MC
realizations) after destriping. The solid horizontal line is the theoretical
white noise spectrum. All noise spectra refer to the modelled full set of 24
{\it PLANCK} LFI 100~GHz radiometers and nearly full sky coverage. }
 \label{clnoise}
\end{figure}

In this study we do not address the question of estimating the
noise from the data.  Thus, whereas in a more realistic simulation
the noise power spectrum to be used for the noise MC simulations
would be estimated from the data as discussed above, we have used
the same underlying noise power spectrum $P(f)$ for the noise MC
simulations as was used for the signal+noise TOD from which we are
estimating the $\widehat{C}_\ell$. This was a mixture of white and
$1/f$ (slope = $-1$) noise with knee frequency $f_k = 0.1$~Hz. The
nominal rms (root-mean-square) level per integration time ($t =
1/f_{\mathrm{sample}} = 1/108.3$~s) is $\sim 4.8$ mK
(thermodynamic temperature scale) for one 100 GHz LFI radiometer.
The full set of 24 radiometers at 100~GHz was modelled by reducing
this number by a factor of $\sqrt{24}$. We used an SDE (Stochastic
Differential Equation)\footnote{SDE is one of the two methods in
the {\it PLANCK} Level S pipeline for producing simulated
instrument noise. The method was implemented by B.~Wandelt and
K.~G\'orski and modified by E.~Keih\"{a}nen.} algorithm to
generate the TODs of instrumental noise. It is unrealistic to
extend the $1/f$ noise spectrum behaviour to very low frequencies.
Therefore we have set a minimum frequency $f_{\rm min} = 4\times
10^{-6}$~Hz ($\sim (70~ \hbox{h})^{-1}$) below which the noise
spectrum becomes flat. We used $N_{\rm MC}^{\rm(n)} = 100$ noise
TOD realizations to produce the estimate of
$\langle\widetilde{N_\ell}\rangle$. According to \citet{BAL02},
already 22 MC realizations produce a good estimate.

If the instrument noise appears to be non-stationary, SDE can be
applied to generate the overall TOD stream in a piecewise manner
where each piece is assumed to be stationary with its own specific
set of noise model parameter values. The accuracy of the
determination of the baseline magnitudes in the destriping is not
dependent on whether the noise is stationary throughout the TOD or
only piecewise stationary.

Fig.~\ref{clnoise} shows the angular power spectrum of the noise map (no
destriping) compared with the angular power spectrum after the destriping. Two
things are evident. Firstly, without destriping there is a distinct blob
structure which is caused by the geometry of the scanning. The amplitudes of
the blobs depend on the value of $f_k$. The second thing is the significant
capacity of the destriping in removing the blobs and making the spectrum
approach the spectrum of the white noise. The spectrum deviates from the
spectrum of the white noise at low $\ell$. This is an indication that some
residual $1/f$ noise remains after the baseline subtraction.

\section{Scanning Strategy and Window Function}
\label{sec:scanning_strategy}

The order in which the samples of the TOD hit the pixels of the
map is determined by the scanning strategy.  Therefore any
artefacts produced in the maps due to the map-making method can be
expected to be affected by the scanning strategy. In this study we
used the nominal {\it PLANCK} scanning strategy.

In this strategy the spin axis of the satellite follows the ecliptic plane. The
spin axis is kept anti-solar by repointing it by $2.5$~arcmin every hour. The
spacecraft rotates around the spin axis at a nominal rate of 1 rpm. During one
hour {\it PLANCK} scans the same circle on the sky 60 times. The scanning
pattern used in this study corresponds to the 100 GHz LFI detector number 10
(location $(\theta,\phi) = (3\fdg737,126\fdg228)$ in the focal plane image).
The angle between the satellite spin axis and the optical axis of the telescope
is $85\degr$.
We assumed no spin axis precession. The
antenna beam pattern has been modelled by a symmetric Gaussian beam with
nominal resolution of FWHM = 10~arcmin (Full Width Half Maximum).

Since we assume idealized satellite motion, where the 60 circles between
repointings fall exactly on each other, sample by sample, we can immediately
average these circles into a single ring.  Thus our simulated TOD consists of
5040 scanning rings, corresponding to 7 months of measurement time, with 6498
samples on each ring, corresponding to a sampling frequency of
$f_\mathrm{sample}=108.3$ Hz.

The sky coverage of the applied scanning strategy was
$f_\mathrm{sky} = 0.985$, leaving areas close to the poles
unobserved. This case is called ``nearly full sky'' in this paper.
In addition, we considered a case called ``galactic cut'', where
the galactic region ($|b|\leq 20^\circ$) is cut out from the maps,
leaving $f_\mathrm{sky} = 0.646$. The numbers of hits in the
output map ($N_\mathrm{side} = 512$) are shown in Fig.~\ref{hits}.
The polar regions with no hits and the galactic cut (lower map)
are clearly visible.  The average number of hits per pixel was
$10.6$ in the observed part of the sky (``nearly full sky'' case).

\begin{figure}
\begin{center}
\end{center}
\caption{The maps with numbers of hits per pixel for the nearly
full sky ({\bf top}) and for the galactic cut ({\bf bottom})
cases. The maps are in ecliptic coordinate system. Note that the
scale is $\log_{10}(\rm{n_{hit}})$, where $\rm{n_{hit}}$ is the
number of hits in a pixel.} \label{hits}
\end{figure}

The window functions $W(\mathbf{n})$ applied in the nearly full sky and in the
galactic cut cases were top hat window maps, where each unobserved or cut out
pixel was assigned a value zero and for the rest of the pixels value one was
assigned. The resolution of the window map was the same as the resolution of
our output map. The angular power spectra ($W_\ell$) of these window maps were
determined with the {\sc Anafast} code of the HEALPix package. With the aid of
the angular spectra the kernel matrices were produced (see Eq. (\ref{kernel})).

\section{Filter Function and Signal Bias}
\label{sec:filter_function}

Often some kind of filtering is applied to the data to reduce the
effect of low-frequency noise. The filtering is applied to the TOD
or to the maps. In our case the TOD is ``filtered'' in a sense
that a baseline for each scan circle is subtracted. The magnitudes
of the baselines are mainly determined by the $1/f$ noise.
However, as described in Section~\ref{sec:destri} the baseline
magnitudes can be influenced by the CMB signal as well and
therefore the CMB signal is affected by the baseline subtraction.
This effect on the angular power spectrum has to be accounted for.
The filter function $F_\ell$ was introduced in Eq. (\ref{clgen})
as one possible means to model this effect. The filter function
will also include the power spectrum distortion caused by any
other effects inherent in the map making method.

We determined the filter function $F_\ell$ in the following way.
Starting from a theoretical input CMB angular power spectrum
$C_\ell^{\rm th}$ we produced $N_{\rm MC}^{\rm (s)} = 450$ pure
signal (i.e. no noise) full sky map realizations with
$N_\mathrm{side} = 1024$ by using the {\sc Synfast} code of the
HEALPix package. These maps were then observed using the nominal
{\it PLANCK} scanning strategy\footnote{Simulation Software is
part of the Level S of the {\it PLANCK} DPCs and is available for
the {\it PLANCK} collaboration at
http://planck.mpa-garching.mpg.de} (see
Section~\ref{sec:scanning_strategy}).

The resulting TODs were destriped and coadded into maps.
The angular resolution of the crossing point search and the
resolution of the output maps were both 7~arcmin (corresponding to
$N_{\rm side}=512$). The pseudo angular power spectra
($\widetilde{C}_\ell$) were then extracted from the output maps.
The $\widetilde{C}_\ell$ were also determined from maps with no
baselines removed (no destriping). By comparing the filter
functions with and without destriping we can distinguish from
other effects those filter function contributions which are due to
the baseline removal.

The average of these MC realizations, $\langle\widetilde{C}_\ell
\rangle_\mathrm{MC}$, was then used in place of
$\langle\widetilde{C}_\ell \rangle$ in Eq. (\ref{clgen}) and
$F_\ell$ was solved from this equation (with
$\langle\widetilde{N}_\ell \rangle = 0$).

\begin{figure}
\begin{center}
\includegraphics[width=8cm,height=4.8cm]{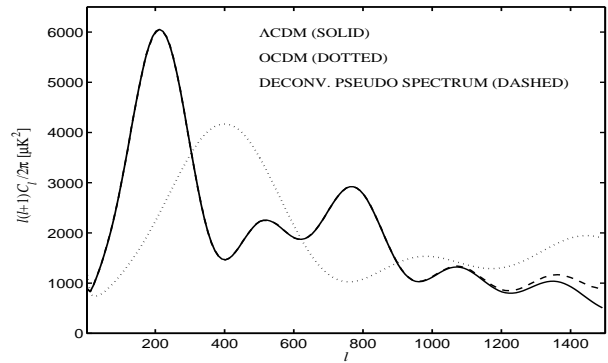}
\end{center}
\caption{The input CMB power spectra $C_\ell^{th}$ of the
$\Lambda$CDM model (solid curve) and the OCDM model (dotted
curve). The ensemble average ($N_{\rm MC}^{\rm (s)} = 450$) of the
deconvolved pseudo power spectrum
from signal-only simulations of the $\Lambda$CDM model is shown as
well (dashed curve). The spectrum was derived from the destriped
maps and the sky coverage was nearly full sky. The filter function
is the ratio of the deconvolved power spectrum (dashed curve) and
the input power spectrum (solid curve).}
 \label{spectra}
\end{figure}

In order to find out whether the filter function $F_\ell$ depends
on the sky signal we considered two widely different cosmological
models: a $\Lambda$CDM (cosmological constant + Cold Dark Matter)
model and an open CDM model (OCDM). Their power spectra
$C_\ell^\mathrm{th}$ are depicted in Fig.~\ref{spectra}. They were
computed from the respective cosmological models by using the {\sc
Cmbfast}\footnote{http://www.cmbfast.org} code (see
\citealt{SEL96} and references therein). As an example, the
ensemble average of a deconvolved pseudo power spectrum
$\sum_{\ell'}{M_{\ell \ell'}^{-1}\langle \widetilde{C}_{\ell'}
\rangle_{MC}} / B^2_\ell$ in the case of nearly full sky and
$\Lambda$CDM is shown in Fig.~\ref{spectra}. The filter function
is a ratio of this spectrum and the theoretical power spectrum
$C_\ell^\mathrm{th}$ (compare to Eq.\ (\ref{clestimate})).

\begin{figure}
\begin{center}
\includegraphics[width=8cm,height=4.8cm]{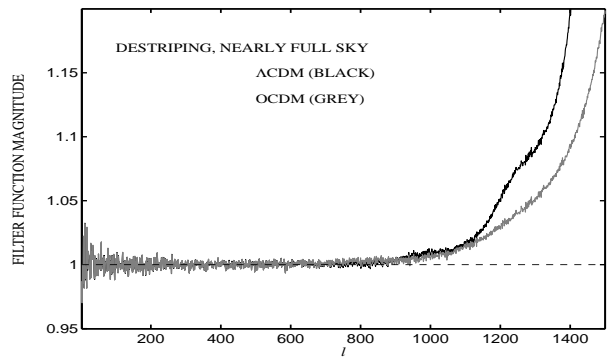}
\end{center}
\caption{Filter functions $F_\ell$ for two widely different input
CMB power spectra $C_\ell^{th}$. The sky coverage is nearly full
sky. The resulting filter functions after the baselines have been
removed from the TODs (destriping) are shown.
The filter functions are results of $N_{\rm MC}^{\rm (s)} = 450$
MC realizations. The filter functions of destriping and no
destriping are nearly identical.} \label{filter}
\end{figure}

\begin{figure}
\begin{center}
\includegraphics[width=8cm,height=3.0cm]{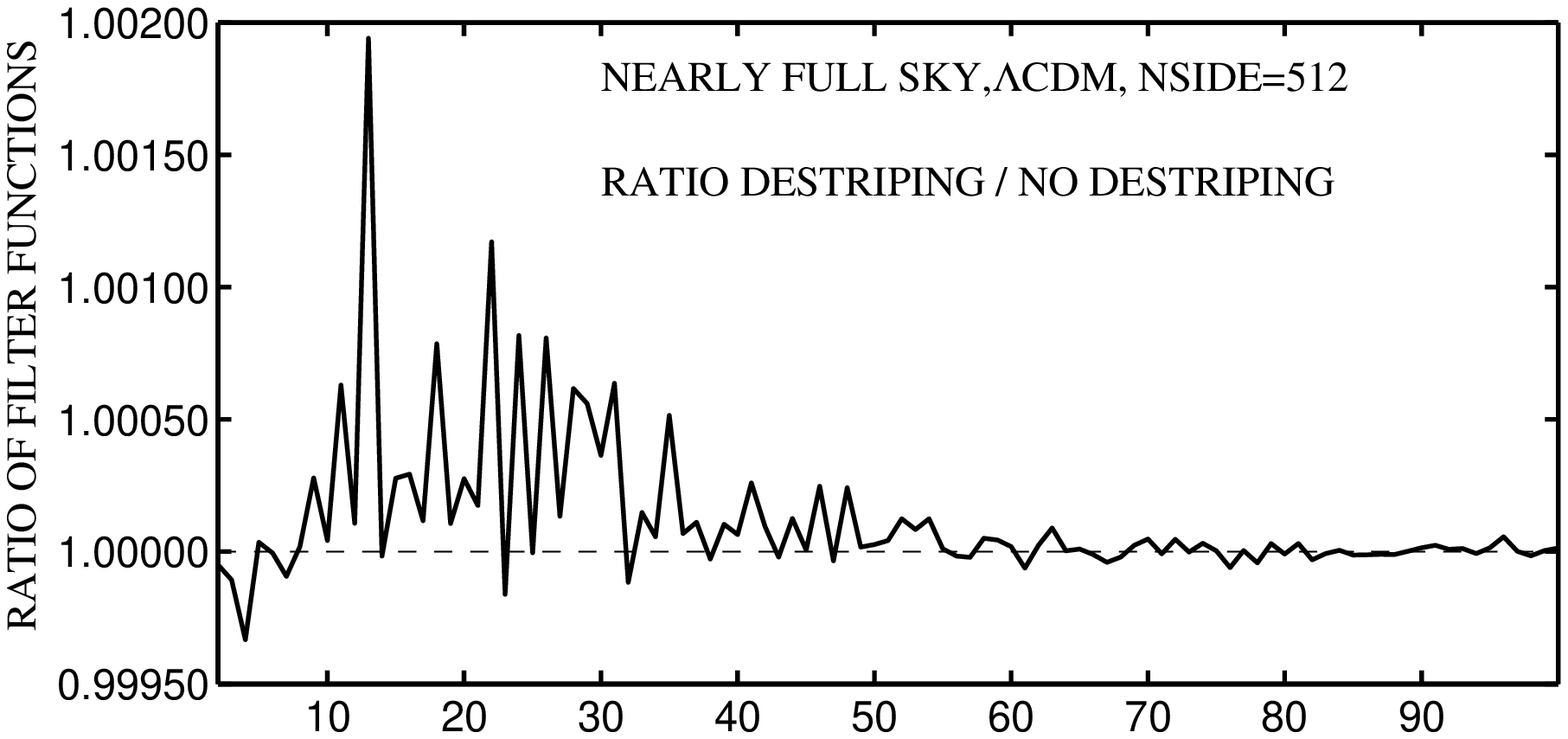}
\includegraphics[width=8cm,height=3.0cm]{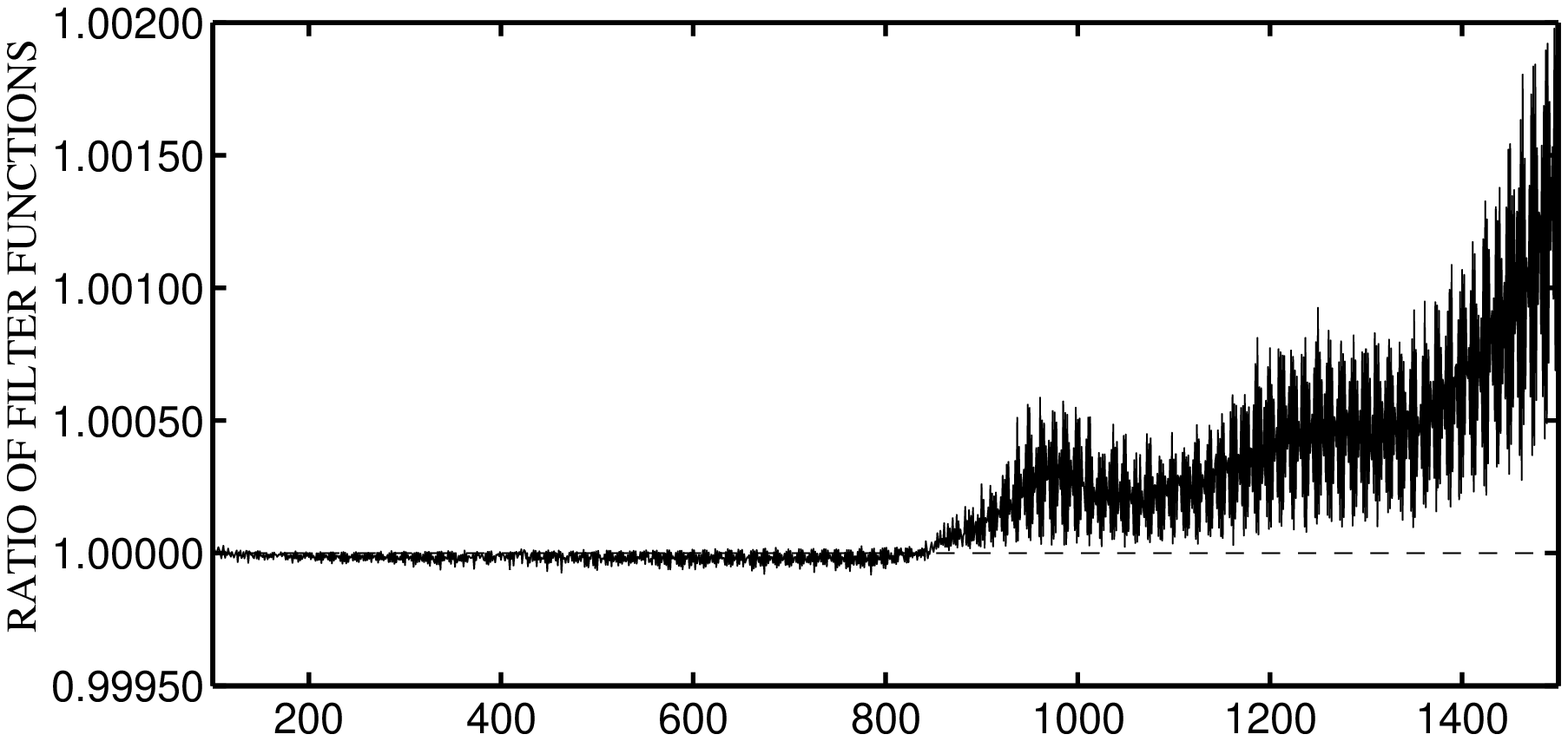}
\includegraphics[width=8cm,height=3.6cm]{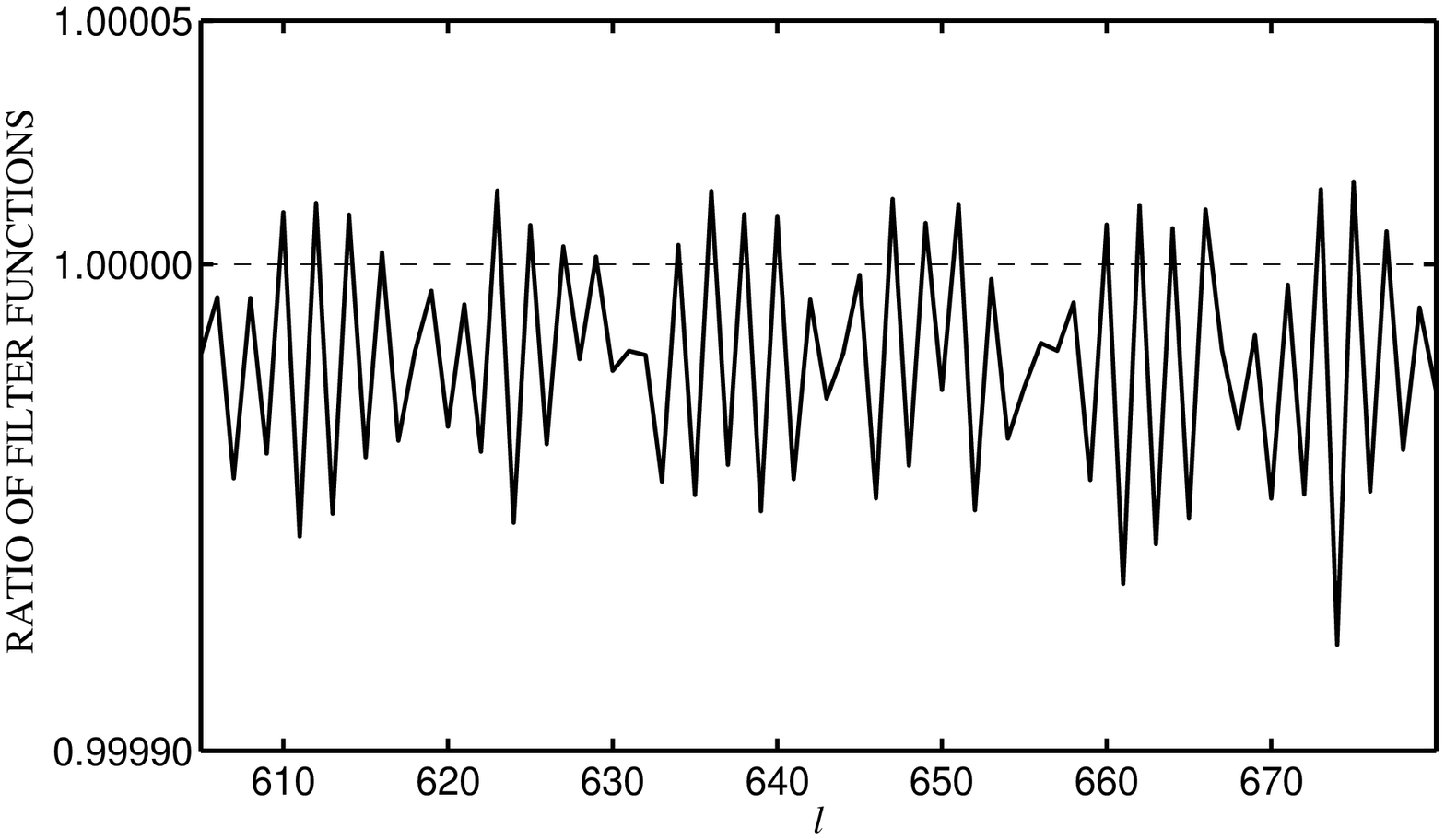}
\end{center}
\caption{The ratio between filter functions with and without
destriping. Nearly full sky coverage and $\Lambda$CDM cosmological
model were applied. Note that different panels show the ratio at
different ranges of $\ell$.}
 \label{fs2}
\end{figure}

The resulting filter functions with destriping and for the nearly
full sky are shown in Fig.~\ref{filter}.  (The galactic cut case
looks similar, just somewhat noisier.) In order to see in more
detail the effect of destriping the ratio between the filter
functions with and without destriping were derived. They are shown
in Fig.~\ref{fs2}. A number of things are evident (see
Figs.~\ref{filter} and~\ref{fs2}):
\begin{itemize}
  \item[(1)]
The filter function stays close to one up to $\ell \simeq$ 700.
The deviations from one are completely dominated by remaining
cosmic variance (reduced here by $(N_\mathrm{MC}^{(s)})^{-1/2}$
due to averaging over $N_\mathrm{MC}^{(s)}$ realizations). Thus we
conclude that the effect of our map-making method on the angular
power spectrum is very small for $l < 700$, and we could just use
$1$ for the filter function, i.e., no filter function is needed
for this part of the spectrum.
 \item[(2)]
For $l>700$ the filter function deviates from $1$ significantly.  This
high-$\ell$ tail of the filter function is strongly dependent on
$C_\ell^\mathrm{th}$.
 \item[(3)]
The filter functions with and without destriping are nearly
identical. This shows that the high-$\ell$ tail is caused by some
other effects than the baseline removal. The largest relative
impact due to baseline removal is around 0.2~per cent (see
Fig.~\ref{fs2}). The ratio contains blob structure due to
low-level stripes introduced by the baseline removal (cf.
Fig.~\ref{clnoise}).

\end{itemize}

\begin{figure}
\begin{center}
\includegraphics[width=8cm,height=4.8cm]{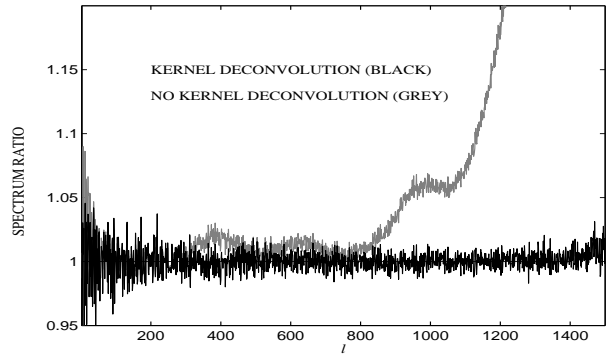}
\end{center}
\caption{CMB power spectra were produced by first generating 100
realizations of the full sky, then cutting out part of the
observations (galactic cut case, see Fig.~\ref{hits}), then taking
the mean of the pseudo power spectra derived from the partial sky
maps and finally deconvolving the mean with beam, pixel weight and
kernel matrix. The ratio between the deconvolved spectrum and the
theoretical input spectrum $C_\ell^{th}$ is shown (black curve).
The ratio without deconvolution with the kernel matrix is
demonstrated as well (grey curve). The cosmological model was
$\Lambda$CDM.
All maps had $N_\mathrm{side} = 512$.}
 \label{kernel_check}
\end{figure}

Before drawing further conclusions we need to investigate which
part of the high-$\ell$ excess power of the deconvolved pseudo
spectrum (see Fig.~\ref{spectra}) is caused by numerical errors
introduced by the procedures we use for generating sky maps,
deriving pseudo spectra from the maps and deconvolving the pseudo
spectra with beam, pixel weight and mode-mode coupling kernel
matrix ($M_{\ell \ell'}$). For that purpose we produced 100
realizations of the full sky CMB maps from the $\Lambda$CDM input
power spectrum, applied our ``galactic cut'' window function, and
determined the pseudo spectra of the resulting maps.
The mean of these pseudo spectra was then deconvolved with the
beam, with the pixel weight (provided by HEALPix package) and with
the kernel matrix. The resulting deconvolved spectrum had a close
resemblance to the theoretical input spectrum $C_\ell^{th}$. Their
ratio is depicted in Fig.~\ref{kernel_check} (black curve). The
ratio stays close to one at all multipoles considered, showing
that the procedures we use for the generation of sky maps and
pseudo spectra and for the deconvolution are not responsible for
the high-$\ell$ tail of the filter function (shown in
Fig.~\ref{filter}).

\begin{figure}
\begin{center}
\includegraphics[width=8cm,height=4.8cm]{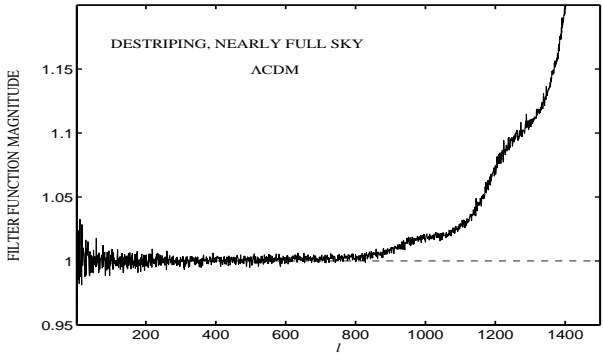}
\end{center}
\caption{Same as Fig.~\ref{filter} but now the sky map
realizations (scanned for TOD) and the output map had higher
resolutions ($N_\mathrm{side} = 2048$ and $N_\mathrm{side} =
1024$, respectively). The filter function is a result of $N_{\rm
MC}^{\rm (s)} = 450$ MC realizations.} \label{filter_1024}
\end{figure}

The dependence of the filter function on $C_\ell^\mathrm{th}$ (see
Fig.~\ref{filter}) shows that it is not a good way to model the
effect of our map-making method. That the effect appears at high
$\ell$ suggests that it might be related to the pixel size used.
Therefore we determined the filter functions also for map resolutions
$N_\mathrm{side} = 2048$ (sky map) and $N_\mathrm{side} = 1024$ (output map).
Results are shown in Fig.~\ref{filter_1024}: a high-$\ell$ tail
still exists, and is in fact somewhat higher than in the earlier
case of $N_\mathrm{side} = 512$ (output map). Thus the filter
function tail cannot be removed simply by increasing the map
resolution. (Note that the $N_\mathrm{side} = 512$ pixel size is
about 7~arcmin and should thus resolve the power spectrum up to
$\ell \sim 180^\circ/7~\hbox{arcmin} \sim 1500$.)

The binning of the detector pointings in pixels has been treated
in time domain by \citet{DOR01}. In a noise-free experiment they
considered a residual between an observed TOD and another TOD
produced by scanning a pixelized map that is naively coadded from
the observed TOD. This residual was called pixelization noise.

We have determined (Appendix~\ref{sec:pointing_distribution}) that
the main effect causing the high-$\ell$ tail of the filter
function is related to the distribution of the detector pointings
on the sky and its relation to the pixelization. Comparing the
filter functions of the $\Lambda$CDM and OCDM cases to the
corresponding $C_{\ell}^\mathrm{th}$ one notices that they appear
to be inversely related. Since the filter function is essentially
a ratio between $\langle \widehat{C}_\ell \rangle$ and
$C_{\ell}^\mathrm{th}$, this suggests that the dominant
contribution
would be better modelled as an offset $S_\ell$, which we call {\em
signal bias},
 \beq
   \langle \widetilde{C}_\ell \rangle = \sum_{\ell'} M_{\ell\ell'}
   B_{\ell'}^2 C_{\ell'}^{th} + S_\ell + \langle\widetilde{N}_\ell \rangle \,.
   \label{tildeT8}
 \eeq
It can be estimated by signal-only MC simulations just like the
filter function. An equation similar to Eq. (\ref{clestimate}) can
be written for the estimate of $C_{\ell}^{th}$,
 \beq
   \widehat{C}_\ell = \frac{\sum_{\ell'}{M_{\ell \ell'}^{-1}
   \left(\widetilde{C}_{\ell'} - S_{\ell'} - \langle \widetilde{N}_{\ell'}
   \rangle \right)}}{B^2_\ell}.
 \label{clestimate_sb}
 \eeq

The average pseudo power spectrum of the MC realizations,
$\langle\widetilde{C}_\ell \rangle_\mathrm{MC}$, can be used in
place of $\langle\widetilde{C}_\ell \rangle$ in Eq.
(\ref{tildeT8}) and $S_\ell^{\mathrm{MC}}$ can be solved from this
equation, with $\langle\widetilde{N}_\ell \rangle = 0$.

Further details of the signal bias are derived in
Appendix~\ref{sec:pointing_distribution}. It is shown there that the signal
bias estimates are rather similar even though the cosmological models are
widely different.

The signal bias is reduced when data from several detectors are
combined, since that increases the density of detector pointings.
We find in Appendix~\ref{sec:pointing_distribution} that the
signal bias in the region of high $\ell$ is nearly inversely
proportional to the number of detectors. In reality, the detector
pointings from successive scan circles also do not fall exactly on
top of each other. Therefore, in the case of the real {\it PLANCK}
experiment this signal bias may be small enough (when compared to
other systematic effects) to be ignored.

In our idealized simulation, which utilizes only pointings of a
single detector and assumes exact overlap of successive scan
circles within the same pointing period, the signal bias is
larger, and shows up prominently in our results. For comparing the
signal bias to other map-making artefacts we carried out
destriping on a noiseless TOD containing CMB and galactic
foreground signals. In this case the signal bias was significantly
larger (close to two orders of magnitude larger at high $\ell$)
than the residual power spectrum due to the subtracted baselines.

We show in Appendix~\ref{sec:pointing_distribution} that the
signal bias can be reliably estimated by MC simulation and thus
removed. We have done this in what follows.

\section{Analysis Pipeline and Computing Time}
\label{sec:times}

The starting point for our analysis is a TOD with CMB+noise (the ``original
TOD''), and a noise power spectrum $P(f)$.  Our analysis pipeline is composed
of the following steps:

\begin{enumerate}
 \item
A total of $N_{\rm MC}^{(\rm n)}$ pure noise TODs are produced,
using $P(f)$. They are destriped and maps are coadded from the
TODs from which the baselines have been removed.  The angular
power spectra of the output maps are produced with the {\sc
Anafast} code. By averaging the obtained spectra an estimate for
$\langle\widetilde{N}_\ell \rangle$ is obtained.
 \item
From the original TOD a destriped map is produced, from which a
power spectrum estimate $\widehat{C}'_\ell$ is obtained, using
Eq.\ (\ref{clestimate_sb}) with $S_\ell = 0$.
 \item
Random generation of $N_{\rm MC}^{(\rm s)}$ pure CMB sky
realizations with the {\sc Synfast} code. The estimate
$\widehat{C}'_\ell$ is used as an input power spectrum. In order
to produce a TOD, the sky is observed according to the selected
scanning strategy. TODs are destriped and projected back onto the
sky.  The maps are input to {\sc Anafast} for the determination of
their angular power spectra. These spectra are used for
determining the signal bias $S_\ell^{\mathrm{MC}}$.
 \item
The power spectrum estimate $\widehat{C}_\ell$ is obtained, using Eq.\
(\ref{clestimate_sb}) with $S_\ell^{\mathrm{MC}}$.
\end{enumerate}

At this point we have our final power spectrum estimate $\widehat{C}_\ell$.  To
obtain an estimate for the error bars and the covariance matrix, further
simulations are required.  In the case of a real experiment, these simulations
would be based on the $\widehat{C}_\ell$ just obtained.  However, in this study
we have also available the  $C_\ell^\mathrm{th}$ used to produce the original
TOD.  The emphasis of this study is to reveal the actual error bars of our
method, rather than to simulate the estimation of these error bars in a real
experiment.  Thus we use this $C_\ell^\mathrm{th}$ as the input spectrum to
generate an ensemble of MC estimates $\widehat{C}_\ell$.

\begin{enumerate}
\setcounter{enumi}{4}
 \item
A total of $N_{\rm MC}^{(\rm{s+n})}$ randomly generated
``experimental'' TODs (\ie~with sky signal and noise) are
produced. They are destriped and coadded into a map. Thereafter
the pseudo power spectra $\widetilde{C}_\ell$ are determined.
 \item
The estimated angular power spectra $\widehat{C}_\ell$ are computed using Eq.
(\ref{clestimate_sb}) and binned.
 \item
The $\pm 1\sigma$ error bars on each $\widehat{C}_\ell$ and bin
are evaluated as std (standard deviation) calculated over the MC
realizations.
\end{enumerate}

In our simulations we had $N_{\rm MC}^{(\rm n)} = 100$, $N_{\rm MC}^{(\rm s)} =
450$, and $N_{\rm MC}^{(\rm{s+n})} = 450$. The relatively large number of
signal+noise MC cycles was required to study the statistics of the
$\widehat{C}_\ell$ estimates. The relative rms accuracy of the $\pm 1\sigma$
error bars is $(2N_{\rm MC}^{(\rm{s+n})})^{-1/2}$. With $N_{\rm
MC}^{(\rm{s+n})} = 450$
this implies 3.3~per cent rms accuracy.
Additionally, this number of MC realizations is high enough to
reveal some features of the covariance matrix $\widehat{C}_{\ell\ell'} =
\langle(\widehat{C}_\ell - \langle \widehat{C}_\ell
\rangle)(\widehat{C}_{\ell'} - \langle \widehat{C}_{\ell'} \rangle)\rangle$
(Section~\ref{sec:simul}). However, to obtain estimates for individual
off-diagonal elements of $\widehat{C}_{\ell\ell'}$, or even of the binned
covariance matrix $\widehat{\mathcal{C}}_{bb'} =
\langle(\widehat{\mathcal{C}}_b - \langle \widehat{\mathcal{C}}_b
\rangle)(\widehat{\mathcal{C}}_{b'} - \langle \widehat{\mathcal{C}}_{b'}
\rangle)\rangle$ would have required an even larger $N_{\rm MC}^{(\rm{s+n})}$.
For $N_{\rm MC}^{(\rm{s+n})} = 450$ the off-diagonal elements were still
dominated by MC noise for bins of $\Delta \ell = 10$.

As for the original MASTER approach, our pipeline is well parallelisable in the
sense that each MC cycle can be run in its own CPU. This allows for a linear
scaling with the number of CPUs and with the number of MC cycles performed. In
our pipeline the most CPU time consuming step is the SDE noise generation. In
an AIX IBM pSeries690 machine with Power4 processor running at clock speed of
1.1~GHz it takes $\sim 1$~h on a single processor to produce one noise TOD for
a single detector with length corresponding to 7 months of mission time ($\sim
2\times 10^9$ samples). We have also implemented a parallel version of the SDE
code which brings the required time to generate one noise TOD down to $\sim
9$~min with 14 processors.

Destriping and map coaddition of one 7 month TOD
will take $\sim 50$~s of CPU time in a single processor serial
job.

\section{Simulations and Results}
\label{sec:simul}

To test our joint destriping and MASTER approach in the estimation
of the angular power spectra we considered simulations relevant
for the 100~GHz channels originally planned for the LFI instrument
on-board the {\it PLANCK} satellite. The simulations included CMB
sky and instrument noise realizations. The applied signal+noise
simulation procedure was described in Section~\ref{sec:times}.

The scanning strategy, antenna beam response and the instrument
noise characteristics applied in the simulations were those
described in Sections~\ref{subsec:noisePS} and
~\ref{sec:filter_function}. When the noise TODs were generated the
effect of using the full set of 24 radiometers was modelled as
described in Section~\ref{subsec:noisePS}. When the signal TODs
were generated a single detector scanning was assumed.

\begin{figure}
\begin{center}
\end{center}
\caption{{\bf Top panel}: Output map without destriping. {\bf
Bottom panel}: Output map after destriping.}
 \label{maps}
\end{figure}

\begin{figure}
\begin{center}
\end{center}
\caption{{\bf Top panel}: Noise-only output map without
destriping. {\bf Bottom panel}: Noise-only output map after
destriping.}
 \label{noisemaps}
\end{figure}

\begin{figure}
\begin{center}
\includegraphics[width=8cm,height=4.8cm]{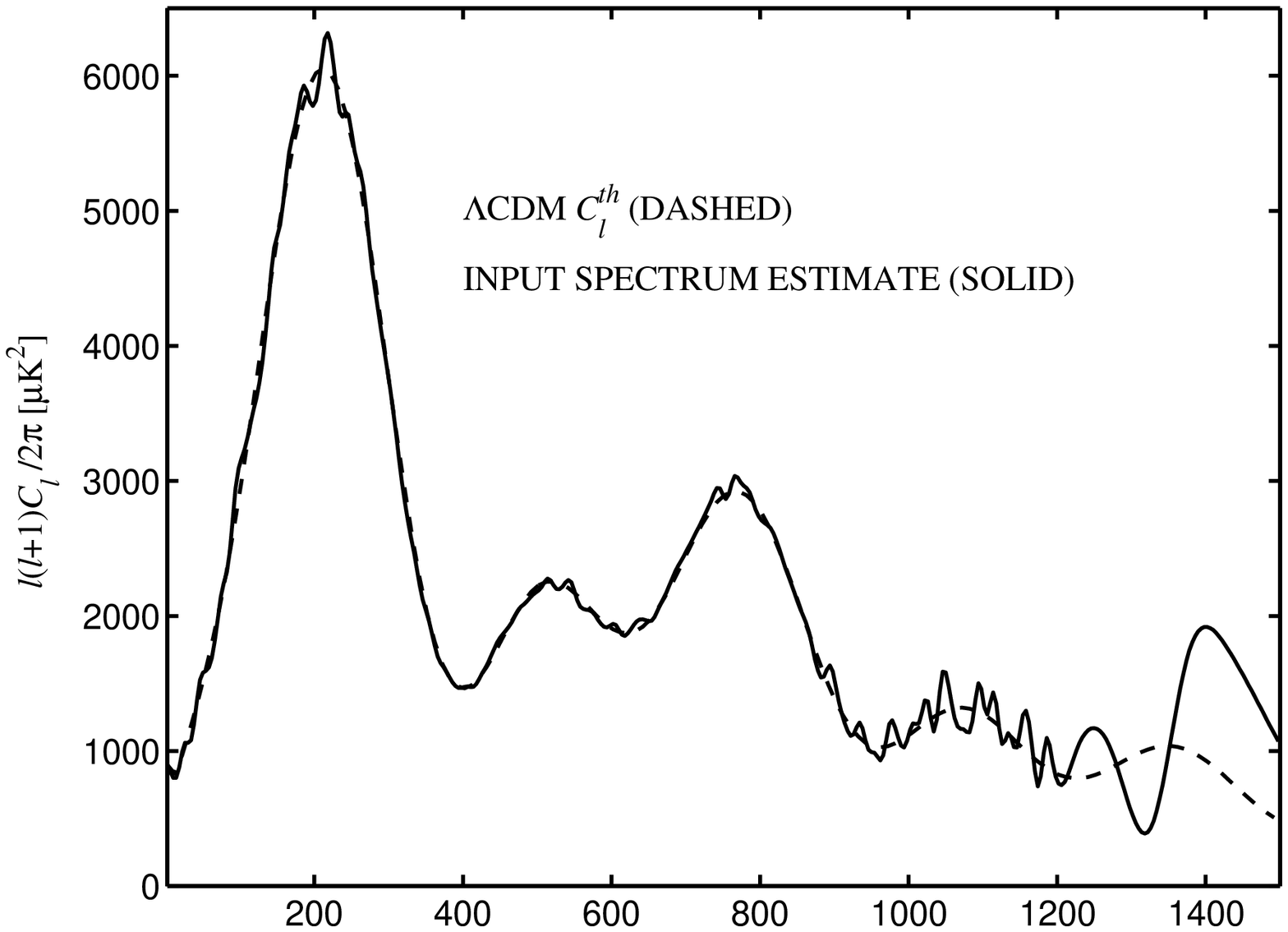}
\includegraphics[width=8cm,height=4.8cm]{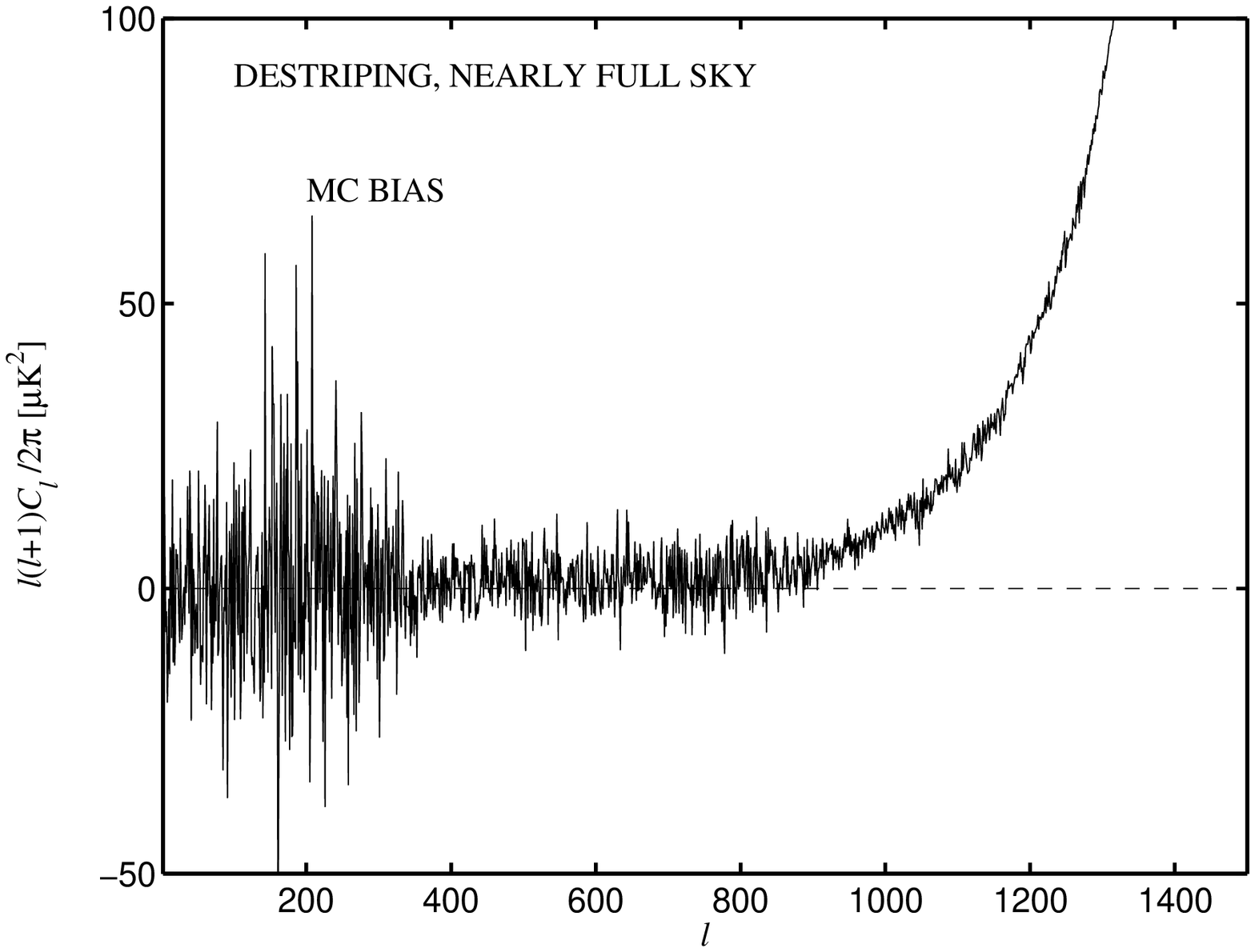}
\end{center}
\caption{{\bf Top panel} shows the noisy power spectrum estimate
(solid curve) used as an input spectrum to the signal-only MC to
determine the estimate for the signal bias. The true
$C_\ell^\mathrm{th}$ is shown as well ($\Lambda$CDM, dashed
curve). {\bf Bottom panel} shows the obtained signal bias
estimate. The sky coverage is nearly full sky.} \label{sb_vs_ff_1}
\end{figure}

\begin{figure}
\begin{center}
\includegraphics[width=8cm,height=6cm]{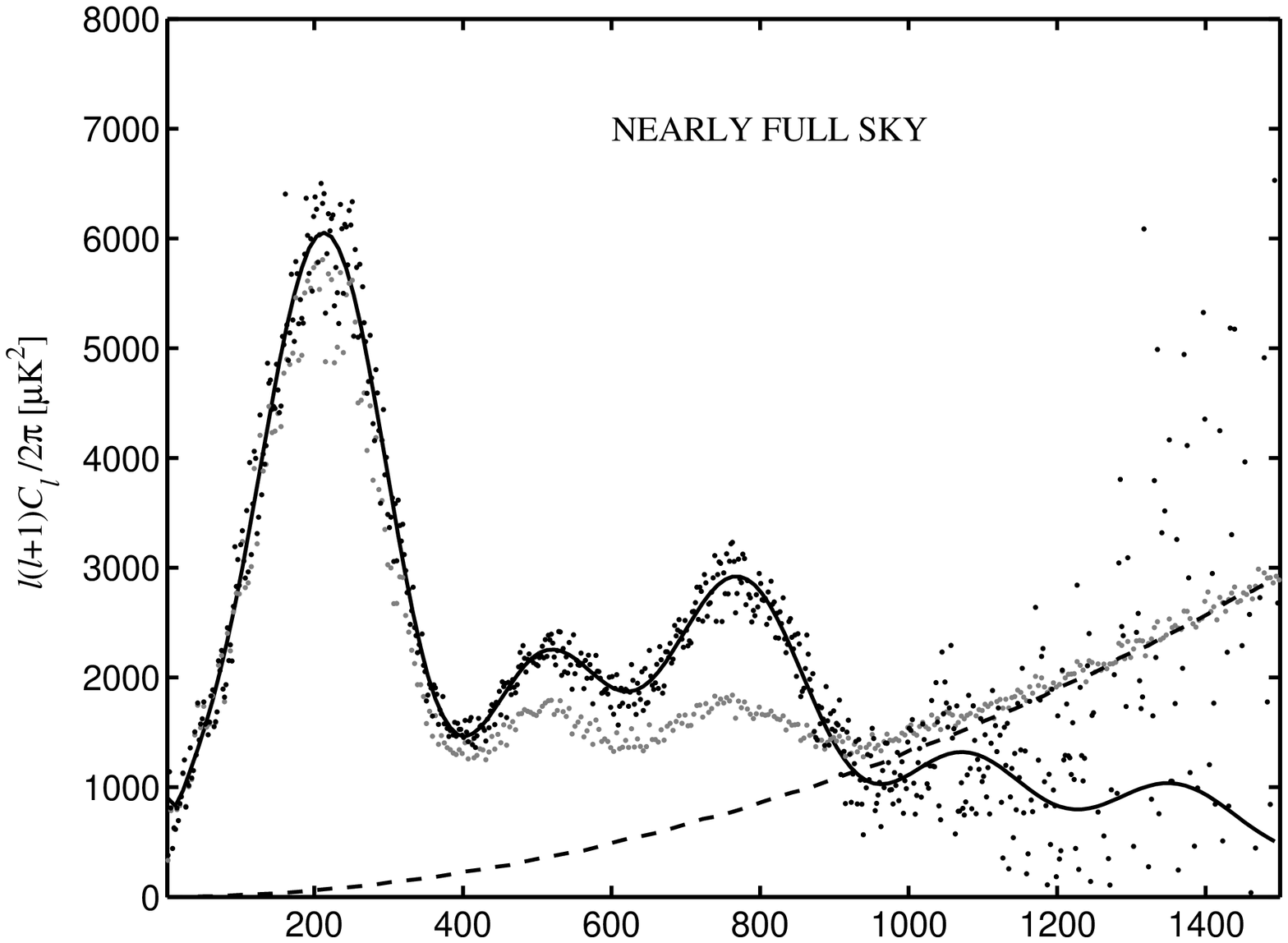}\\
\includegraphics[width=8cm,height=6cm]{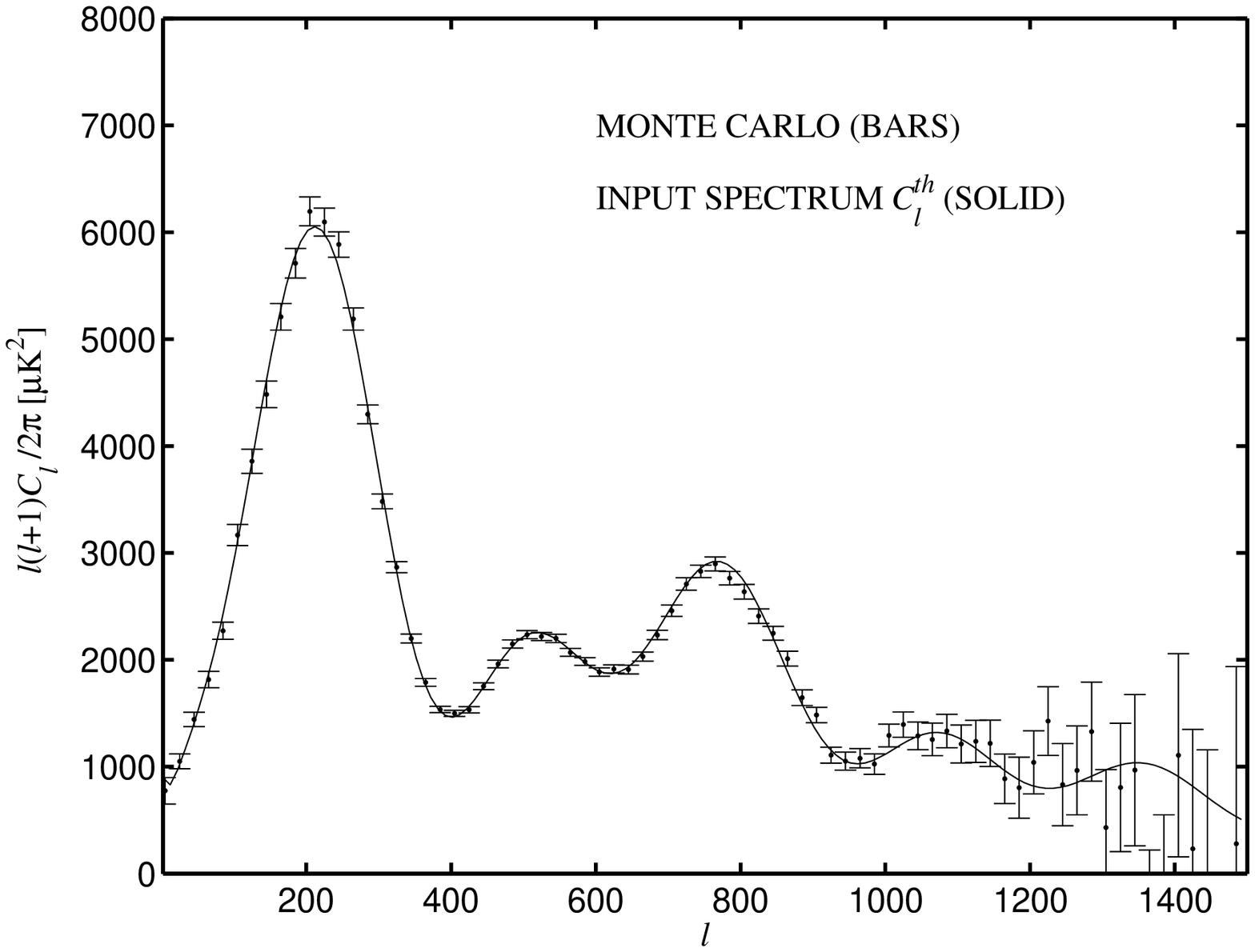}
\end{center}
\caption{CMB power spectrum estimated by using the destriping technique and
MASTER approach for the 100~GHz channels of {\it PLANCK} LFI. The use of 24
radiometers has been modelled by dividing the rms white noise level of a single
detector with a factor $\sqrt{24}$. Nearly full sky has been observed. {\bf Top
panel} shows the power spectra obtained for a single sky and noise realization:
pseudo power spectrum $\widetilde{C}_\ell$ - grey points, the estimated power
spectrum of the noise - dashed line, the estimate of the CMB power spectrum
$\widehat{C}_\ell$ - black points, and the input power spectrum of the true sky
$C_\ell^{th}$ - solid line. {\bf Bottom panel} shows the estimate (midpoints of
the error bars) of the CMB power spectrum in bins of 10. The error bars are
derived from $N_{\rm MC}^{(\rm{s+n})} = 450$ MC realizations of signal and
noise. The solid line is the input power spectrum. The magnitudes of the error
bars are $\pm 1\sigma$. For clarity, only the odd bins are shown.} \label{full}
\end{figure}

\begin{figure}
\begin{center}
\includegraphics[width=8cm,height=6cm]{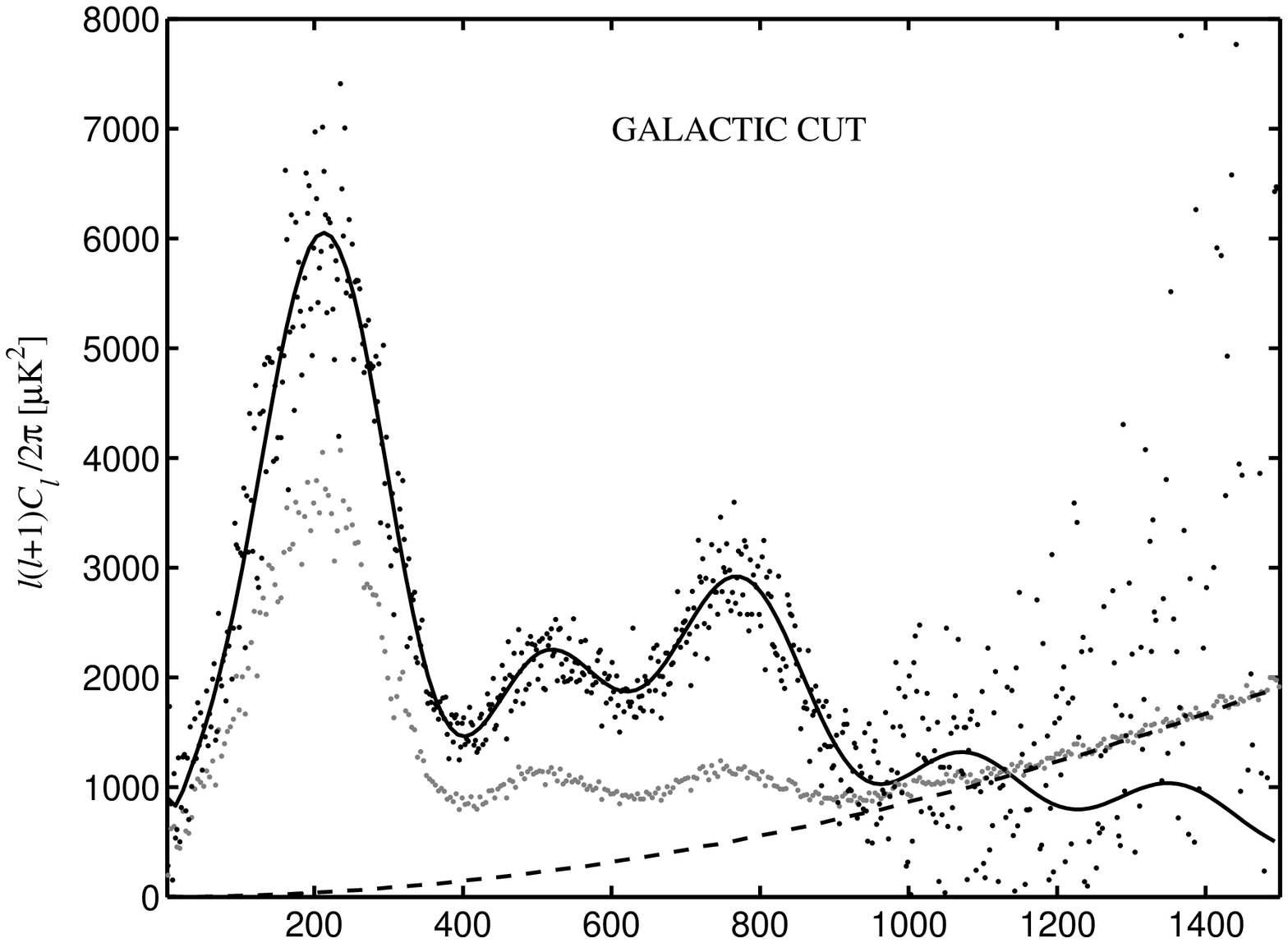}
\includegraphics[width=8cm,height=6cm]{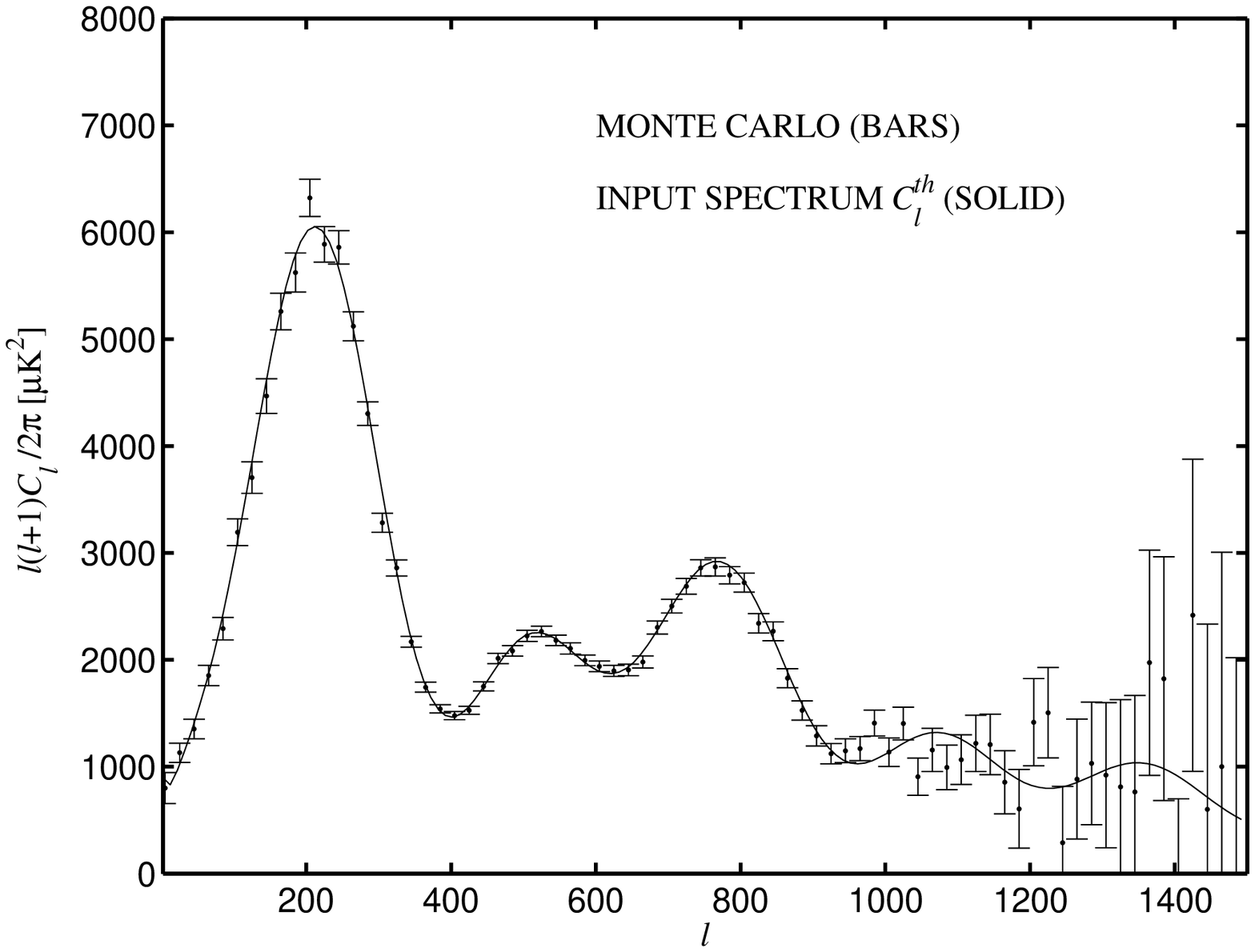}
\caption{Same as Fig.~\ref{full}, except that the galactic region
with $|b|\leq 20^\circ$ has been cut out from the observations.}
 \label{mask}
\end{center}
\end{figure}

The angular power spectrum $C_\ell^{th}$ of the $\Lambda$CDM
cosmological model was used as the input spectrum in the
signal+noise simulations. The applied input spectrum is depicted
in Fig.~\ref{spectra}.

Fig.~\ref{maps} shows a map produced from a simulated CMB+noise
TOD, both before and after destriping.  Fig.~\ref{noisemaps} shows
the same for a pure noise TOD.

\subsection{Power Spectrum Estimate}

The power spectrum estimates $\widehat{C}_\ell$ were solved from
Eq. (\ref{clestimate_sb}). In a real CMB experiment we will not
have the true angular power spectrum $C_\ell^{th}$ available for
generating the signal bias estimate $S_\ell^{\rm{MC}}$. Instead a
noisy power spectrum estimate can be utilized as described in
Section~\ref{sec:times}. The power spectrum estimate and the
signal bias for nearly full sky are shown in
Fig.~\ref{sb_vs_ff_1}. The signal bias for the galactic cut can be
obtained similarly. The applied noise bias $\langle
\widetilde{N}_\ell \rangle$ is depicted in the bottom panel of
Fig.~\ref{clnoise} (nearly full sky). The kernel matrices
($M_{\ell\ell'}$) derived in Section~\ref{sec:filter_function}
were applied.

The resulting angular power spectrum estimates and the
corresponding $\pm 1\sigma$ error bars are shown in
Figs.~\ref{full} and \ref{mask} for the nearly full sky and for
the cut sky cases, respectively.
Individual $\widehat{C}_\ell$ are dominated by the instrument noise for $\ell >
1000$.

\begin{figure}
\begin{center}
\includegraphics[width=8cm,height=5cm]{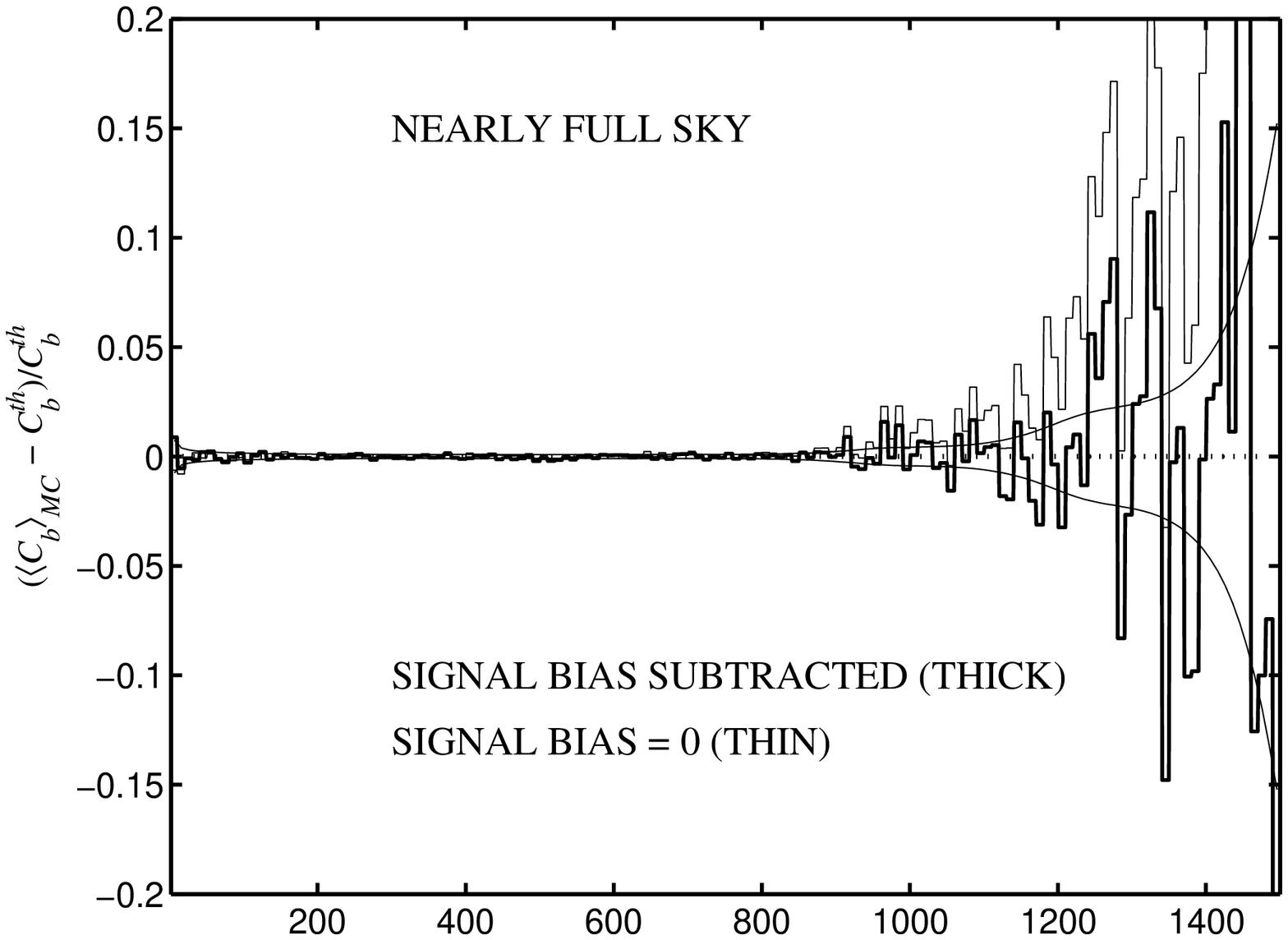}
\includegraphics[width=8cm,height=5cm]{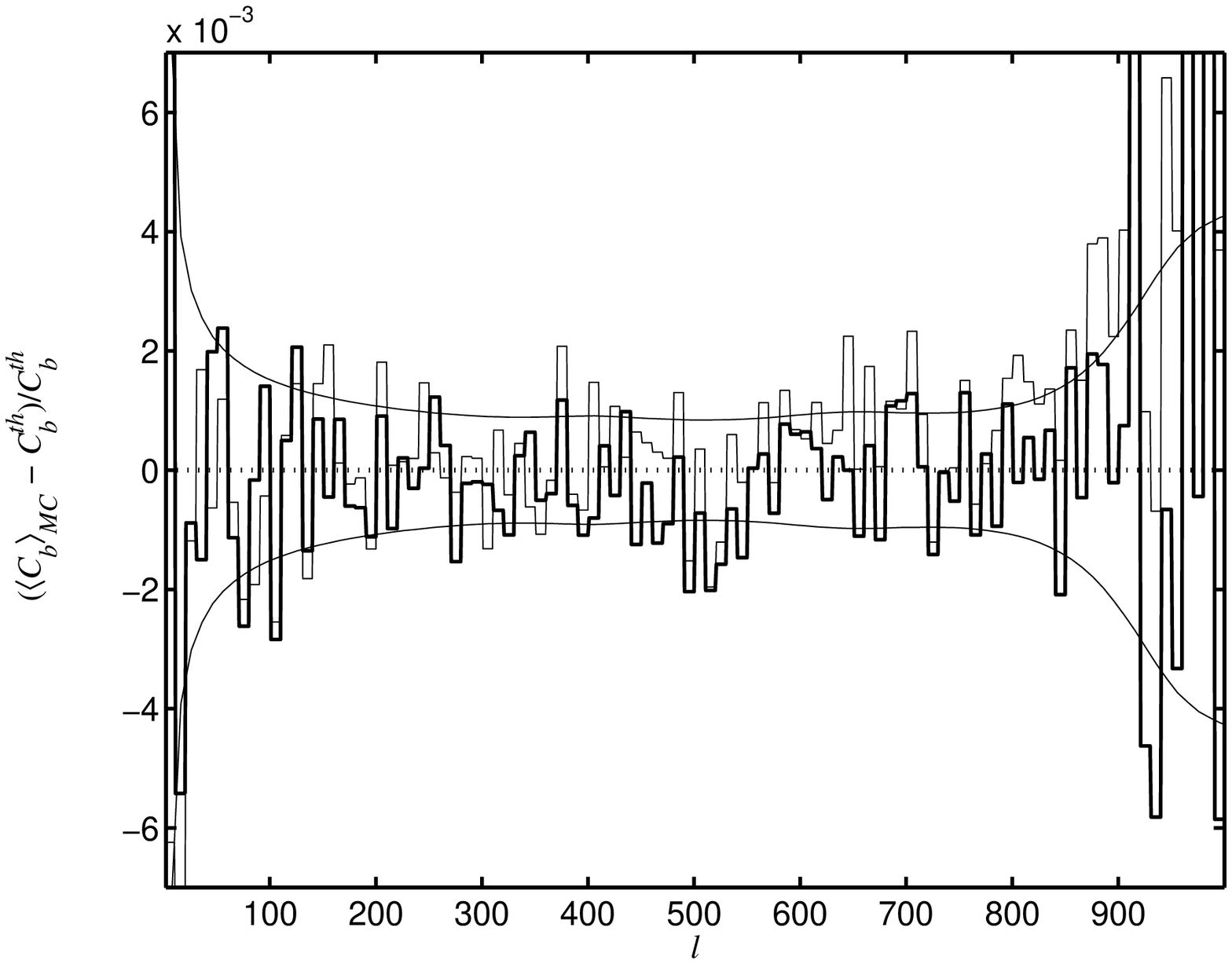}
\end{center}
\caption{In order to reveal the possible bias in our method, we
show the relative differences (thicker step curve) between the
mean binned power spectrum estimate $\langle
\widehat{\mathcal{C}}_b \rangle_{\rm MC}$ (from $N_{\rm
MC}^{(\rm{s+n})} = 450$ MC realizations of signal and noise) and
the binned input power spectrum $\mathcal{C}_b^{\rm th}$
representing the underlying theoretical spectrum. The y axis
quantity is $(\langle\widehat{\mathcal{C}}_b\rangle_{\rm MC} -
\mathcal{C}_b^{\rm th})/\mathcal{C}_b^{\rm th}$. The results of
the nearly full sky are shown.
The symmetric smooth curves show the $\pm \sigma_o$ limits, where
$\sigma_o = \Delta_{\rm ref}
\widehat{\mathcal{C}}_b/(\mathcal{C}_b^{th} (N_{\rm
MC}^{(\rm{s+n})})^{1/2})$ is the ``expected'' standard deviation
from MC noise (the reference value from Eq.~(\ref{stdcb})). No
obvious bias is seen.  The thinner step curve shows the situation
without accounting for the signal bias. In this case there is a
clear positive bias for larger $\ell$. }
 \label{meanreldiff_b10}
\end{figure}

\begin{figure}
\begin{center}
\includegraphics[width=8cm,height=5cm]{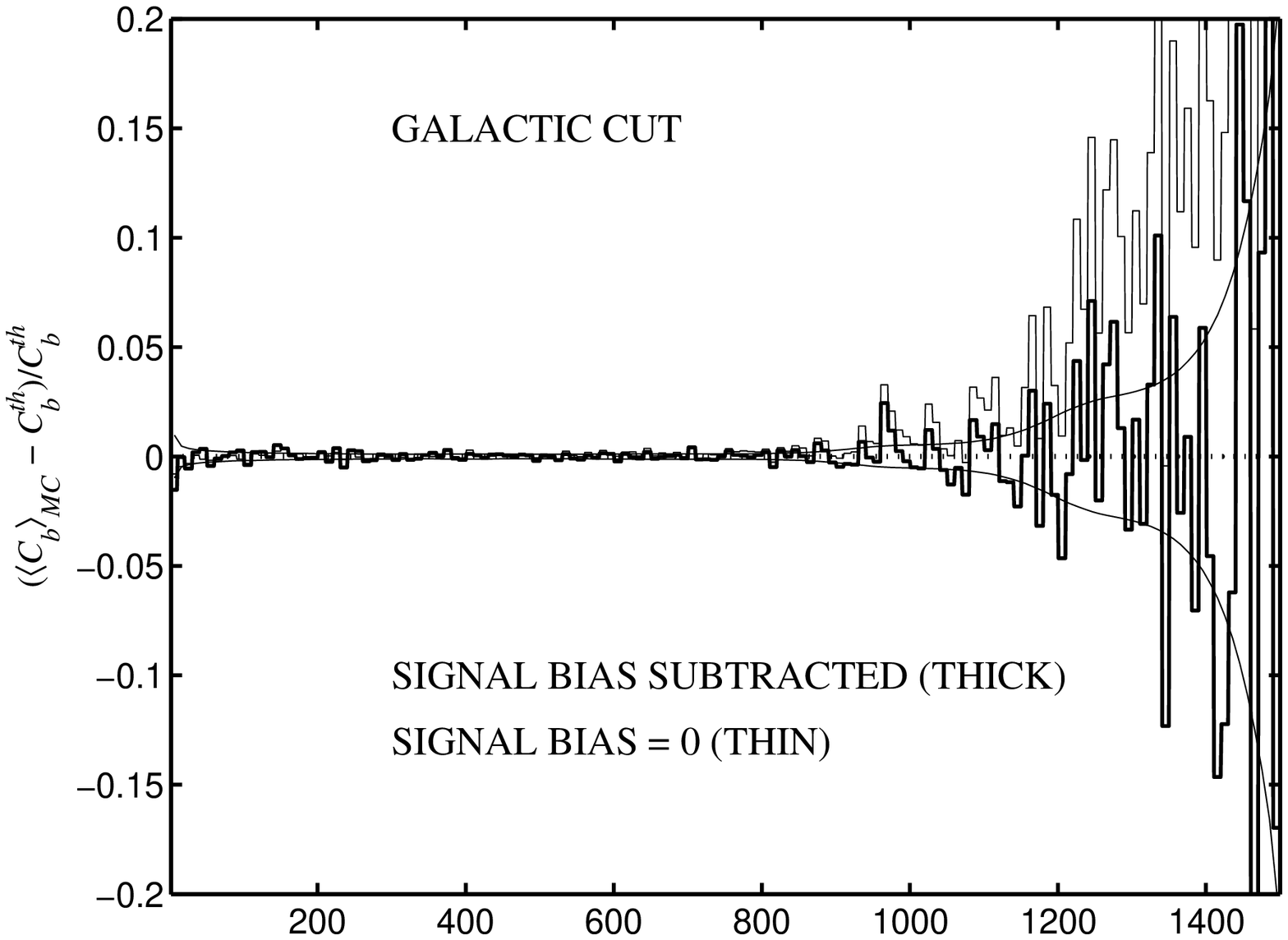}
\includegraphics[width=8cm,height=5cm]{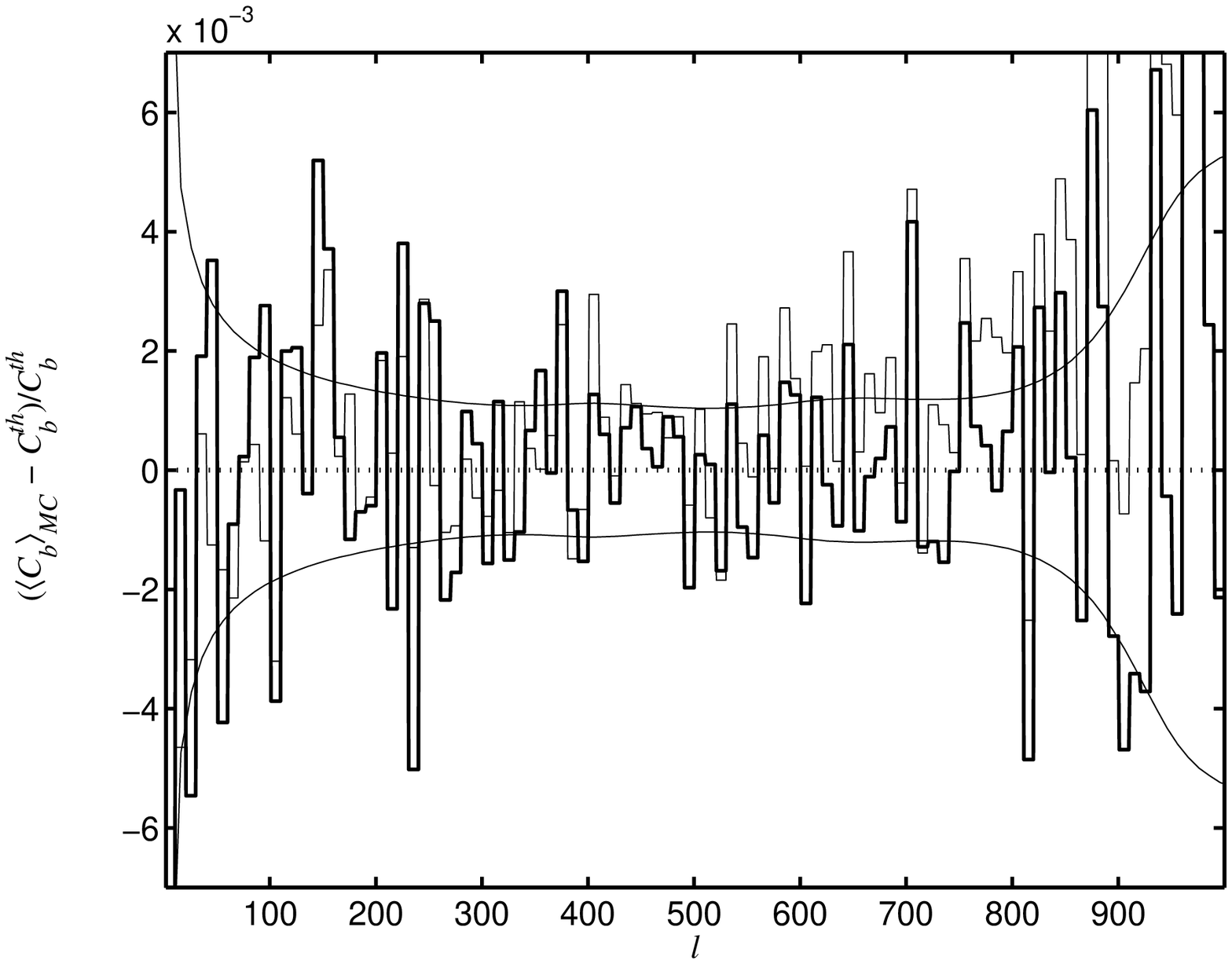}
\end{center}
\caption{Same as Fig.~\ref{meanreldiff_b10} but now the galactic
region is cut out from the output maps.}
 \label{meanreldiff_gb10}
\end{figure}

The bottom panels of Figs.~\ref{full} and \ref{mask} show the
estimates
 \beq
   \widehat{\mathcal{C}}_b = \sum_{\ell \in
   b}{\frac{\ell(\ell + 1)\widehat{C}_\ell}{2\pi\Delta\ell}},
 \label{cb}
 \eeq
of the CMB angular power spectrum in bins $b$ of $\Delta \ell =
10$. The error bars (bottom panels) were determined from the MC
variance of the binned power spectrum estimates. The good match
between the estimated and the input power spectra shows that the
combined destriping and MASTER approach produces power spectrum
estimates with good accuracy.

In order to examine the possible bias in our angular power
spectrum estimates the mean
$\langle\widehat{\mathcal{C}}_b\rangle_{\rm MC}$ of the
$\widehat{\mathcal{C}}_b$ spectra available from our MC
realizations was determined. The relative differences
$(\langle\widehat{\mathcal{C}}_b\rangle_{\rm MC} -
\mathcal{C}_b^{\rm th})/\mathcal{C}_b^{\rm th}$ are shown in
Figs.~\ref{meanreldiff_b10} and \ref{meanreldiff_gb10}.

The signal bias and the power spectrum estimates will be correlated if the same
seed values are used when the random sky realizations are generated in the
signal-only and in the signal+noise simulations.
This could lead to too optimistic results. We intentionally used different seed
values between the signal-only and the signal+noise simulations.

It is evident from Figs.~\ref{meanreldiff_b10} and
\ref{meanreldiff_gb10} that no systematic bias can be observed and
the variation from bin-to-bin is mainly caused by the residual
cosmic variance and the instrument noise that remain after the
averaging.
If any bias exists
its relative level is at most of the order of 0.1~per cent for
$\ell<800$.

If we had not accounted for the signal bias our method would have
produced biased estimates.  To see how large is the effect we did
the same analysis, keeping $S_\ell = 0$.  This case is shown in
Figs.~\ref{meanreldiff_b10} and \ref{meanreldiff_gb10} as well. A
clear bias at high $\ell$ ($\ell > 1000$) can be detected.


\subsection{Covariance Matrix and Statistics}
\label{subsec:covariance}

\begin{figure}
\begin{center}
\includegraphics[width=8cm,height=10cm]{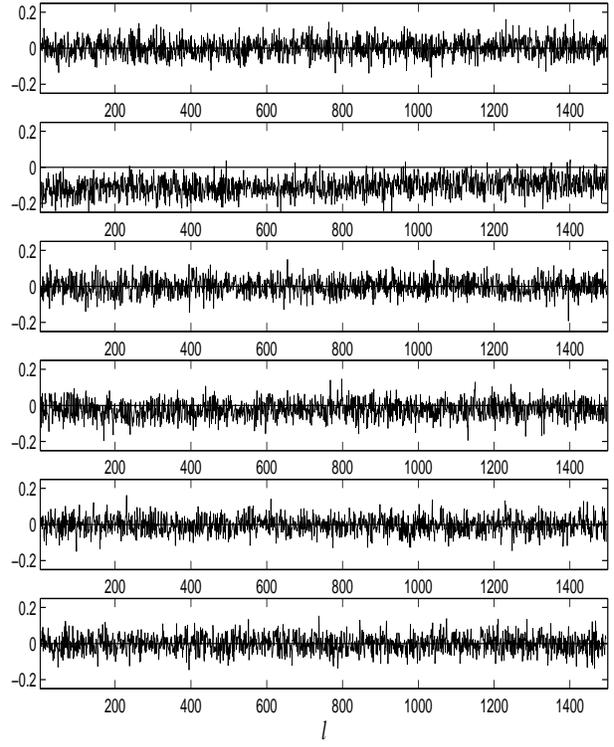}
\end{center}
\caption{The diagonals $\widehat{C}_{\ell,\ell+1} \ldots
\widehat{C}_{\ell,\ell+6}$ of the covariance matrix are shown from top down.
The matrix is normalized to $\widehat{C}_{\ell\ell} = 1$. The
expected std of the MC noise of the off-diagonals is
$(N_{MC})^{-1/2} = 0.047$, where $N_{MC} = 450$. The std of the
diagonals shown in the figure vary in the range 0.046 \ldots
0.050. Note that the levels of the diagonals
$\widehat{C}_{\ell,\ell+2}$ and $\widehat{C}_{\ell,\ell+4}$ are
systematically negative. The values of $\widehat{C}_{\ell,\ell+2}$
seem to increase with increasing $\ell$. The sky coverage is
galactic cut.}
 \label{parallelsgc}
\end{figure}

The covariance matrix of the power spectrum estimates was
determined from the $N_{\rm{MC}} = 450$ signal+noise MC
realizations. The elements of the matrix are
$\widehat{C}_{\ell\ell'} = \langle(\widehat{C}_\ell - \langle
\widehat{C}_\ell \rangle_{\rm{MC}})(\widehat{C}_{\ell'} - \langle
\widehat{C}_{\ell'} \rangle_{\rm{MC}})\rangle_{\rm{MC}}$. The
diagonals $\widehat{C}_{\ell,\ell+1} \ldots
\widehat{C}_{\ell,\ell+6}$ from the upper right triangle are shown
in Fig.~\ref{parallelsgc} (galactic cut). Before plotting, the
covariance matrix was normalized,
 \beq
    \widehat{C}_{\ell\ell'}^{\rm{norm}} =
    \frac{\widehat{C}_{\ell\ell'}}{\sqrt{\widehat{C}_{\ell\ell}}
    \sqrt{\widehat{C}_{\ell'\ell'}}}.
 \label{normalizedCll}
 \eeq
to make all the main diagonal elements equal to one. The $\ell$-to-$\ell$
variation of the off-diagonals is mainly caused by the MC noise.

The second and the fourth diagonals are mostly negative. Their
mean (up to $\ell = 800$) values are -0.116 and -0.0265,
respectively. No level shift can be detected in the sixth or
higher even diagonals. Excluding the diagonals
$\widehat{C}_{\ell,\ell\pm2}$ and $\widehat{C}_{\ell,\ell\pm4}$
the elements of the rest of the covariance matrix were buried
under the MC noise. An analytical model for the covariance matrix
$\widehat{C}_{\ell\ell'}$ of the pseudo-$C_\ell$ estimators exists
\citep{EFS03}. In the CMB dominated region (below $\ell \simeq
800$) it forecasts for the mean levels of the second and the
fourth diagonals -0.113 and -0.0266, respectively. The good match
between the model values and the values obtained from the MC
realizations is notable. The same model gives mean values around
$-1\times 10^{-4}$ for the rest of the diagonals shown in
Fig.~\ref{parallelsgc}. No level shifts that would stand out from
the MC noise could be detected in the covariance matrix of the
nearly full sky.


\begin{figure}
\begin{center}
\includegraphics[width=8cm,height=5cm]{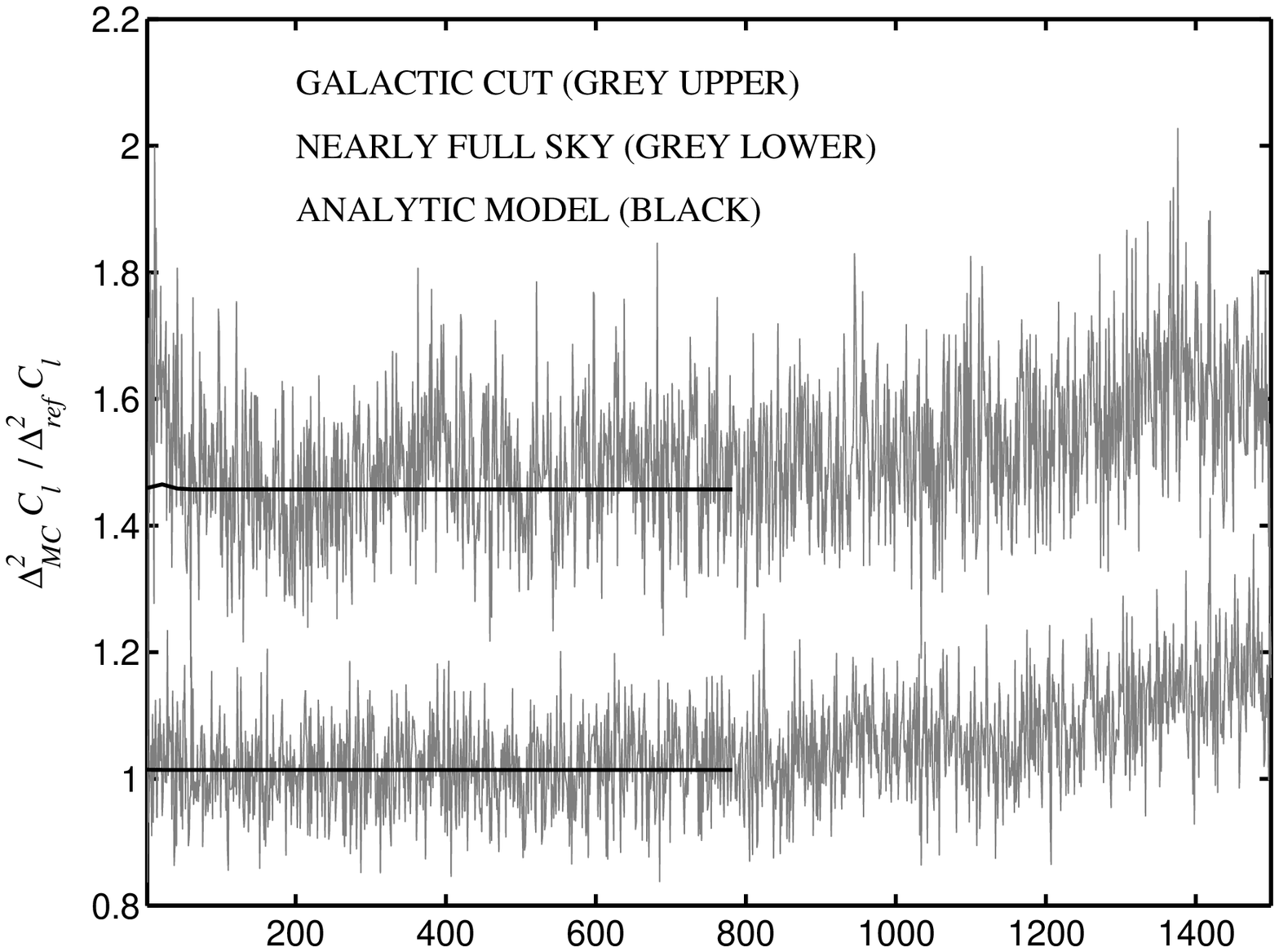}
\includegraphics[width=8cm,height=5cm]{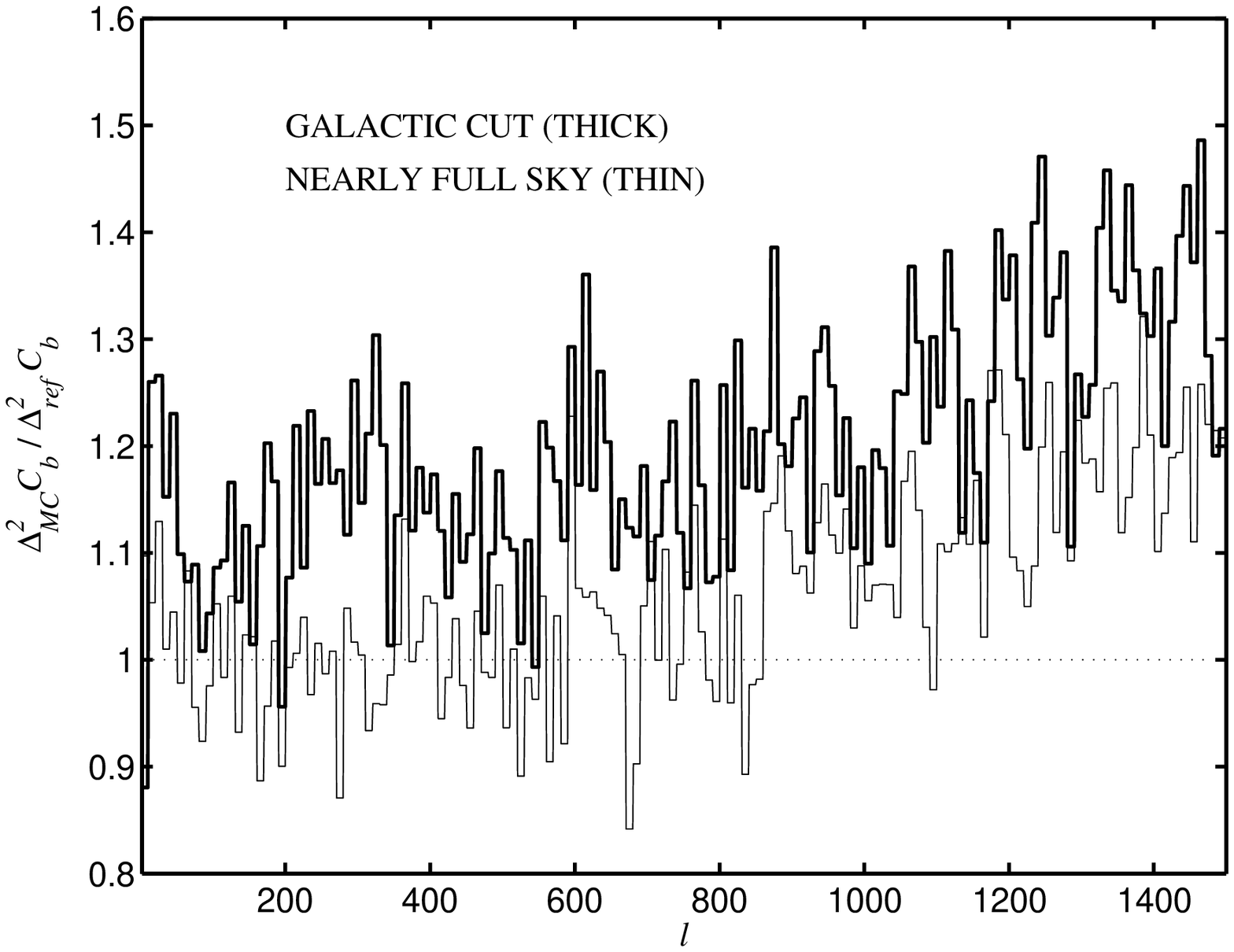}
\end{center}
\caption{The ratios of the variances
$\Delta^2_{\rm{MC}}\widehat{C}_\ell/\Delta^2_{\rm{ref}}\widehat{C}_\ell$
({\bf top panel}) and
$\Delta^2_{\rm{MC}}\widehat{\mathcal{C}}_b/\Delta^2_{\rm{ref}}\widehat{\mathcal{C}}_b$
({\bf bottom panel}). $\Delta^2_{\rm{MC}}$ refers to the MC
variance and $\Delta^2_{\rm{ref}}$ is the reference value defined
in Eqs. (\ref{deltaClcut_sb}) and (\ref{stdcb}). In Eq.
(\ref{deltaClcut_sb}) the quantity
$\langle\widetilde{N}_\ell\rangle_{\rm{MC}}$ obtained from $N_{\rm
MC}^{(\rm n)} = 100$ noise-only MC realizations was used instead
of $\langle\widetilde{N}_\ell\rangle$. The number of signal+noise
MC realizations was 450 when determining the variances.
The horizontal lines in the top panel were derived from the
analytical model of the covariance matrix of the power spectrum
estimates \citep{EFS03}.} \label{khi_ab}
\end{figure}

The theoretical variance for the full-sky case is given by Eq.\
(\ref{thvariance}).  For the cut sky less information is
available, which increases the variance.  The cut sky also
introduces correlations between the estimates $\widehat{C}_\ell$.
The binned values $\widehat{\mathcal{C}}_b$ should be less
correlated. According to a ``rule of thumb'' (for bins
$\widehat{\mathcal{C}}_b$) the variance should be larger
approximately by a factor $1/f_{\rm sky}$
\citep[e.g.][]{SCO94,HOB96}.
Thus we shall use as {\em reference values} for the variances
 \begin{eqnarray}
   \Delta_{\rm ref} \widehat{C}_\ell =
   \sqrt{\frac{2}{(2\ell +1) f_{\rm sky}}} \bigg(C_\ell^{th} +{}
   \nonumber\\ +{}\frac{\sum_{\ell'}{M_{\ell \ell'}^{-1} (S_{\ell'} + \langle
   \widetilde{N}_{\ell'} \rangle})}{B_\ell^2}\bigg)
 \label{deltaClcut_sb}
 \end{eqnarray}
and
 \beq
   \Delta^2_{\rm ref} \widehat{\mathcal{C}}_b = \frac{1}{{\Delta \ell}^2}
   \sum_{\ell \in b}{\left(\frac{\ell(\ell + 1)\Delta_{\rm ref}
   \widehat{C}_\ell}{2\pi}\right)^2}.
 \label{stdcb}
 \eeq

We show the ratios of the MC variances from our simulations to
these reference values,
$\Delta^2_{\rm{MC}}\widehat{C}_\ell/\Delta^2_{\rm{ref}}\widehat{C}_\ell$
and
$\Delta^2_{\rm{MC}}\widehat{\mathcal{C}}_b/\Delta^2_{\rm{ref}}\widehat{\mathcal{C}}_b$,
in Fig.~\ref{khi_ab}.

The top panel of Fig.~\ref{khi_ab} shows that the MC variance of
the unbinned power spectrum estimate is larger than its reference
value. Lowering the sky coverage will produce a larger ratio. The
ratio remains relatively flat up to $\ell \approx 1200$. At higher
$\ell$ a positive bump appears. The horizontal solid lines
indicate the expected levels derived from the analytical model of
the covariance matrix $\widehat{C}_{\ell\ell'}$ \citep{EFS03}. The
model is valid in the CMB dominated region. In the nearly full sky
coverage the MC variances follow closely the model variances. In
the galactic cut the model is valid only at high enough $\ell$
(above $\ell \simeq 5$). The MC variances of the galactic cut are
in average slightly larger than the model variances. The mean
relative difference is 1.8~per cent.

The ratios between the MC variances of the binned estimates and
their reference values are shown in the bottom panel of
Fig.~\ref{khi_ab}. Since the estimates at $\ell$ and $\ell+2$ are
anticorrelated, binning them into the same bin reduces the
variance by more than $\Delta\ell$. The less correlated the bins
are, the closer to the reference value (Eq. (\ref{stdcb})) we
expect their variance to be. A smaller sky coverage implies more
correlation and thus a higher excess variance. For our cut sky the
remaining excess variance is 10 \ldots 20~per cent for bins of
$\Delta\ell = 10$ up to $\ell \sim 1000$. For nearly full sky no
excess variance is seen below $\ell = 800$. For larger $\ell$
there is more excess variance. This happens when the spectrum
becomes noise dominated.

\begin{figure}
\begin{center}
\includegraphics[width=8cm,height=10cm]{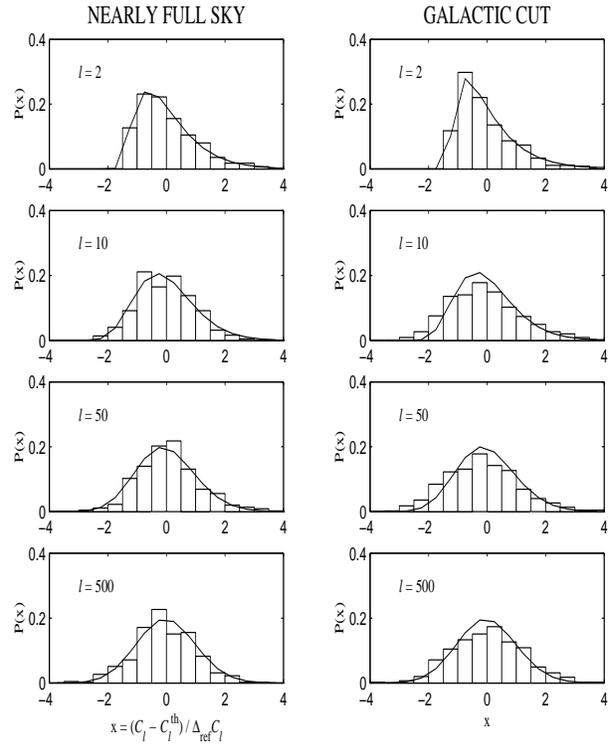}
\end{center}
\caption{PDFs $P(x)$ of $\widehat{C}_\ell$ for some selected multipoles $\ell$.
The bars are histograms derived from 450 signal+noise MC realizations. The
solid curves are the PDFs from the central $\chi_{\nu}^2$ model
with $\nu = (2\ell + 1)f_{\rm sky}$. The std of the model is set to
$\Delta_{\rm{ref}}\widehat{C}_\ell$, which is the reference value of the std of
$\widehat{C}_\ell$ defined in Eq. (\ref{deltaClcut_sb}). In the plot the x axis
quantity of the model PDF is first normalized with
$\Delta_{\rm{ref}}\widehat{C}_\ell$ and then the PDF is shifted along the x
axis so that its expectation value will coincide with $x = 0$.} \label{pdf}
\end{figure}

\begin{figure}
\begin{center}
\includegraphics[width=8cm,height=10cm]{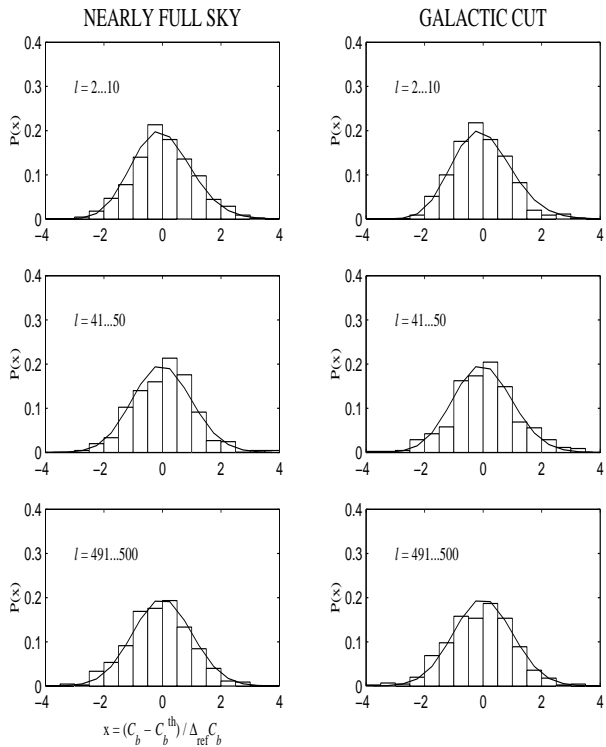}
\end{center}
\caption{Same as Fig.~\ref{pdf} but now the PDFs are for the
binned estimates $\widehat{\mathcal{C}}_b$. In the central
$\chi_{\nu}^2$ model the number of degrees of freedom is set to
$\nu = \sum_{\ell \in b}{(2\ell + 1)f_{\rm sky}}$. The std of the
model is set to $\Delta_{\rm{ref}}\widehat{\mathcal{C}}_b$. The
bin size is $\Delta\ell = 10.$} \label{pdfcb}
\end{figure}

We determined also the statistical distributions of the angular power spectrum
estimates obtained with our $C_\ell$ estimation technique. Examples of the PDFs
(Probability Density Function) of the unbinned ($\widehat{C}_\ell$) and the
binned ($\widehat{\mathcal{C}}_b$) estimates
are shown in Figs.~\ref{pdf} and \ref{pdfcb}. These PDFs were compared to the
PDFs of a central $\chi^2_{\nu}$ model (also shown).
The match between the MC PDFs and the model PDFs is good in the
case of nearly full sky coverage. In the galactic cut the match is
poorer especially in the PDFs of the unbinned estimates. This is
as expected, since the degrees of freedom ($\nu$) in the model
PDFs were determined from the formula $\nu = (2\ell + 1)f_{\rm
sky}$, which is just a "rule of thumb" approximation in the
galactic cut case, and does not correctly take into account the
correlation between nearby $\widehat{C}_\ell$. The obtained PDFs
confirm the earlier finding that the deviation of the MC variance
from its reference value becomes larger when the sky coverage
fraction is lowered (see Fig.~\ref{khi_ab}).

\section{Conclusion}
\label{sec:conclusion}

We have demonstrated that the combination of destriping technique and our
MASTER approach can tackle the extraction of the CMB angular power spectra.
The approach was found to work well yielding accurate estimates of the true
power spectrum of the sky. As a practical example, we considered the 100~GHz
channels of the LFI instrument of the {\it PLANCK} satellite.

Destriping as a map-making method is general in a sense that it requires no
prior information either on the instrument noise properties or on the actual
beam shape. We can expect that even in the case of non-stationary noise we are
able to recover the baseline magnitudes accurately. The MASTER technique
requires knowledge of the noise characteristics. In realistic CMB experiments
the noise parameters need to be estimated from the measured data. However, we
did not address the noise estimation techniques in this study. Thus the correct
instrument noise model was assumed throughout this paper.

We found that the effect of the sky signal on
destriping does not cause a significant distortion on the power spectrum
estimates.  Therefore no filter function is needed to compensate for it.

Instead we discovered an effect related to the clustering of the
detector pointings.  Because of the high stability of the motion
of the {\it PLANCK} satellite the centre points of the
measurements taken during the 1~h period between satellite
spin-axis repointings form tight clusters of 60. In our idealized
simulation these were all assumed to fall on a single point, which
exaggerates the effect somewhat. This leads to a noticeable effect
at high $\ell$ (for $\ell > 700$ in our simulations).  We were
able to account for this effect, and remove it, by introducing the
concept of signal bias in the $C_\ell$ estimate, and determining
it by signal-only MC simulations.  We also derived a reasonably
accurate analytical estimate for the signal bias.

It was shown that the signal bias will be reduced when data from
several detectors are combined, since that increases the density
of detector pointings. In addition, due to satellite spin axis
nutation (amplitudes up to 1.5~arcmin, \citealt{LEE02}) and spin
rate variation the detector pointings from successive scan circles
will not fall exactly on top of each other, in reality. Therefore,
in the case of the real {\it PLANCK} experiment this signal bias
may be small enough to be ignored.

An ensemble of angular power spectrum estimates were produced by
applying our estimation method to the maps of the signal+noise MC
realizations. Our results showed that, after removing the signal
and noise biases, any remaining bias
was buried under the MC variations; if any bias exists its relative level was
not higher than 0.1~per cent.

There was a good match between the $\pm 1\sigma$ error bars
obtained from the MC realizations and derived from the analytical
model of the pseudo-$C_\ell$ estimators \citep{EFS03}. This
indicates that the implementation related performance losses of
our method are small.

\section*{Acknowledgments}
This work was supported by the Academy of Finland Antares Space
Research Programme grant no. 51433. TP wishes to thank the
V\"{a}is\"{a}l\"{a} Foundation for financial support. We thank CSC
(Finland) and NERSC (U.S.A.) for computational resources. The
authors would like to thank J.~Borrill for the support on NERSC
computational facilities. Special thanks go to C.~Cantalupo (from
NERSC) for producing the LFI 70 GHz pointing matrices. We
acknowledge the use of the {\sc Cmbfast} code for the computation
of the theoretical CMB angular power spectra. We gratefully
acknowledge K.~G\'orski and B.~Wandelt for their implementation of
the SDE noise generation method. Some of the results in this paper
have been derived using the HEALPix package \citep{GOR99}.

\appendix
\section{Signal Bias}
\label{sec:pointing_distribution}

In our signal-only simulations (Section~\ref{sec:filter_function})
we noted an effect (at high $\ell$) in the angular power spectra
of the output maps, that was not explained by beam size, pixel
size, or window function, and could not properly be modelled in
terms of a filter function.  We studied this effect in detail, and
conclude that
 \begin{enumerate}
 \item The effect can be modelled as an offset $S_\ell$, which we call {\em signal
bias}.
 \item The effect is due to the resolution by which the sky is sampled by the
 detector pointings.
 \end{enumerate}
Since in the realistic {\it PLANCK} experiment the effect will be less
important than in our idealized simulation, we have relegated the detailed
discussion of this effect in this Appendix.

\begin{figure}
\begin{center}
\includegraphics[width=8cm,height=4.8cm]{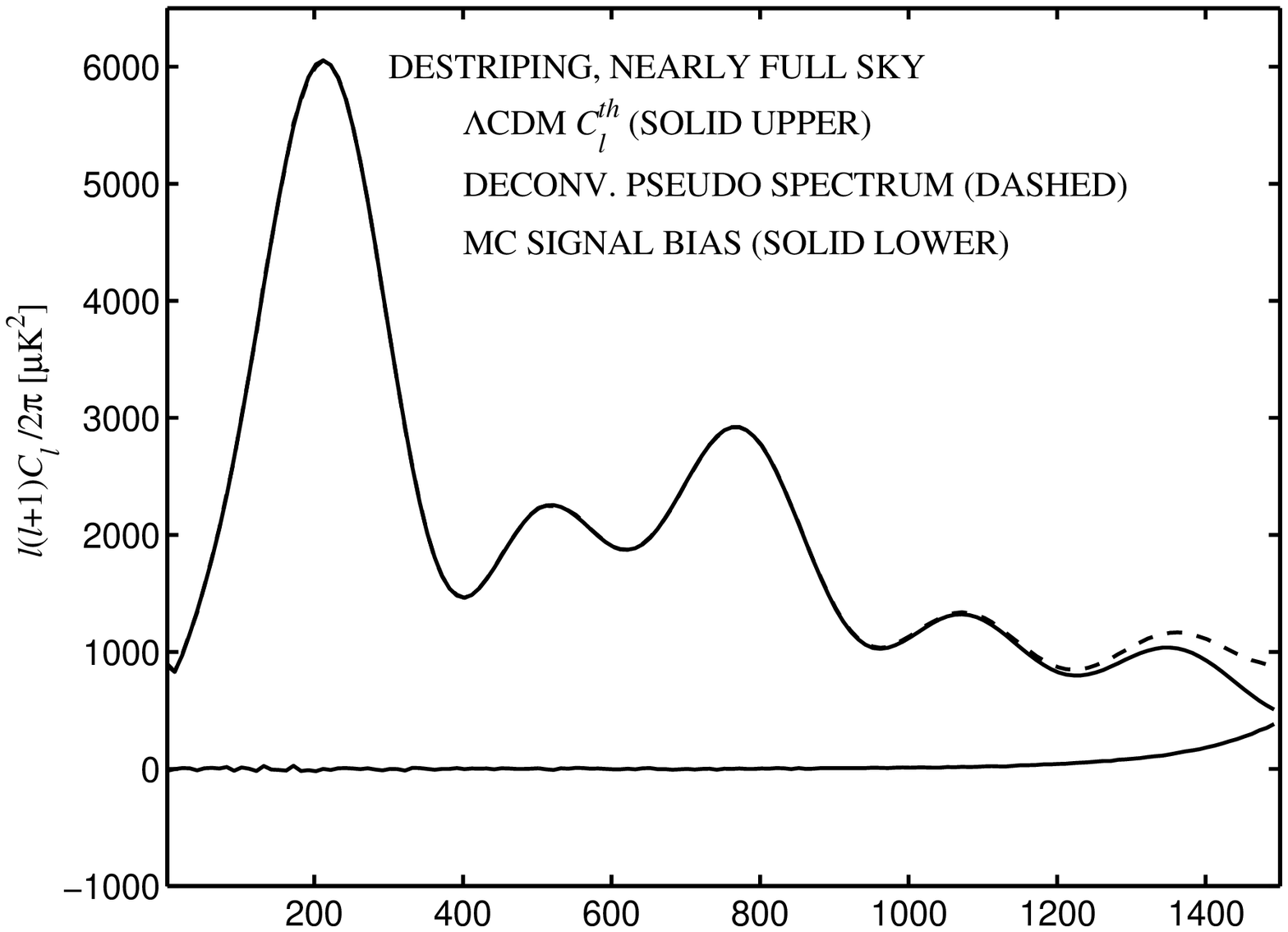}
\includegraphics[width=8cm,height=4.8cm]{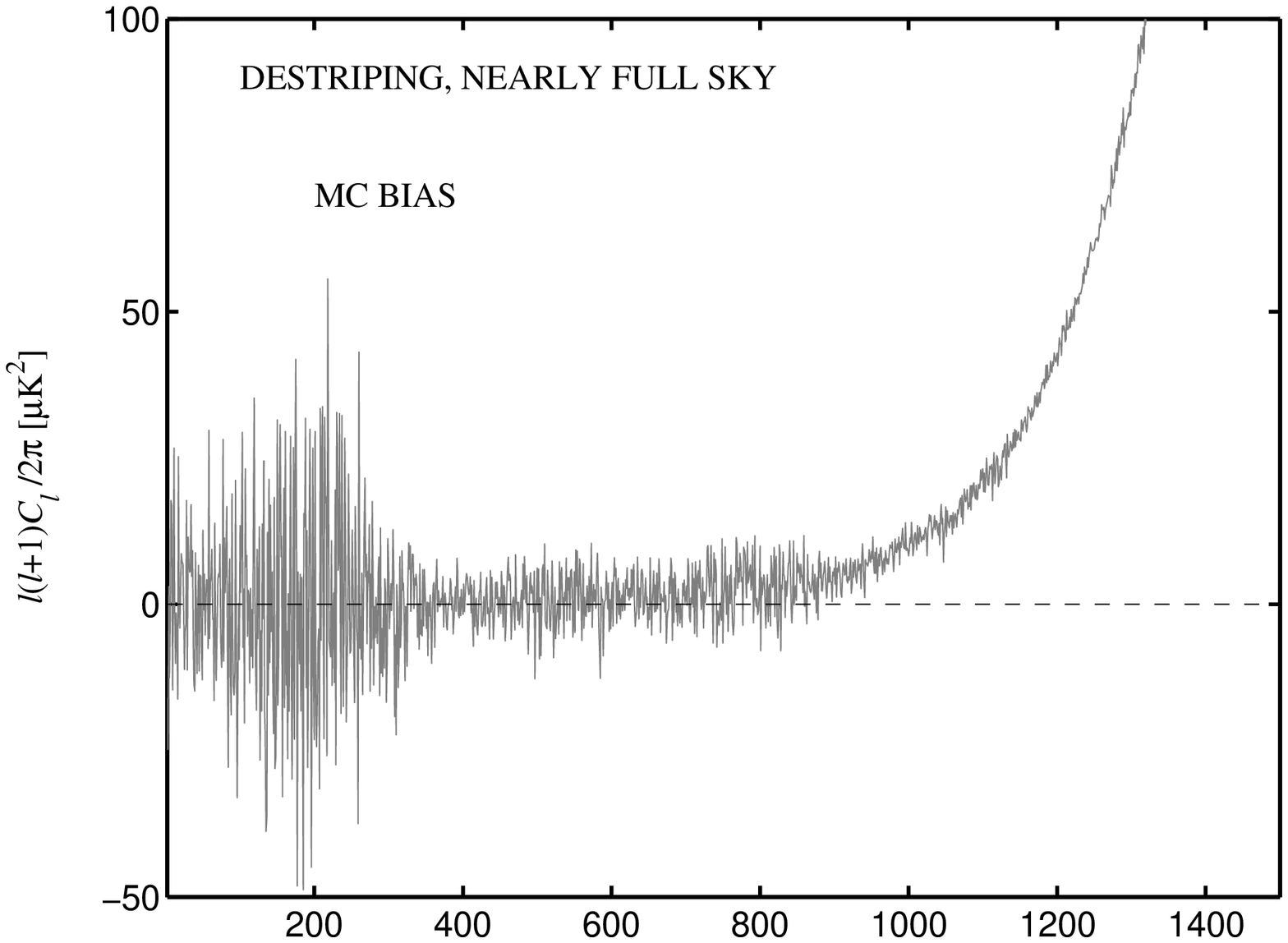}
\includegraphics[width=8cm,height=4.8cm]{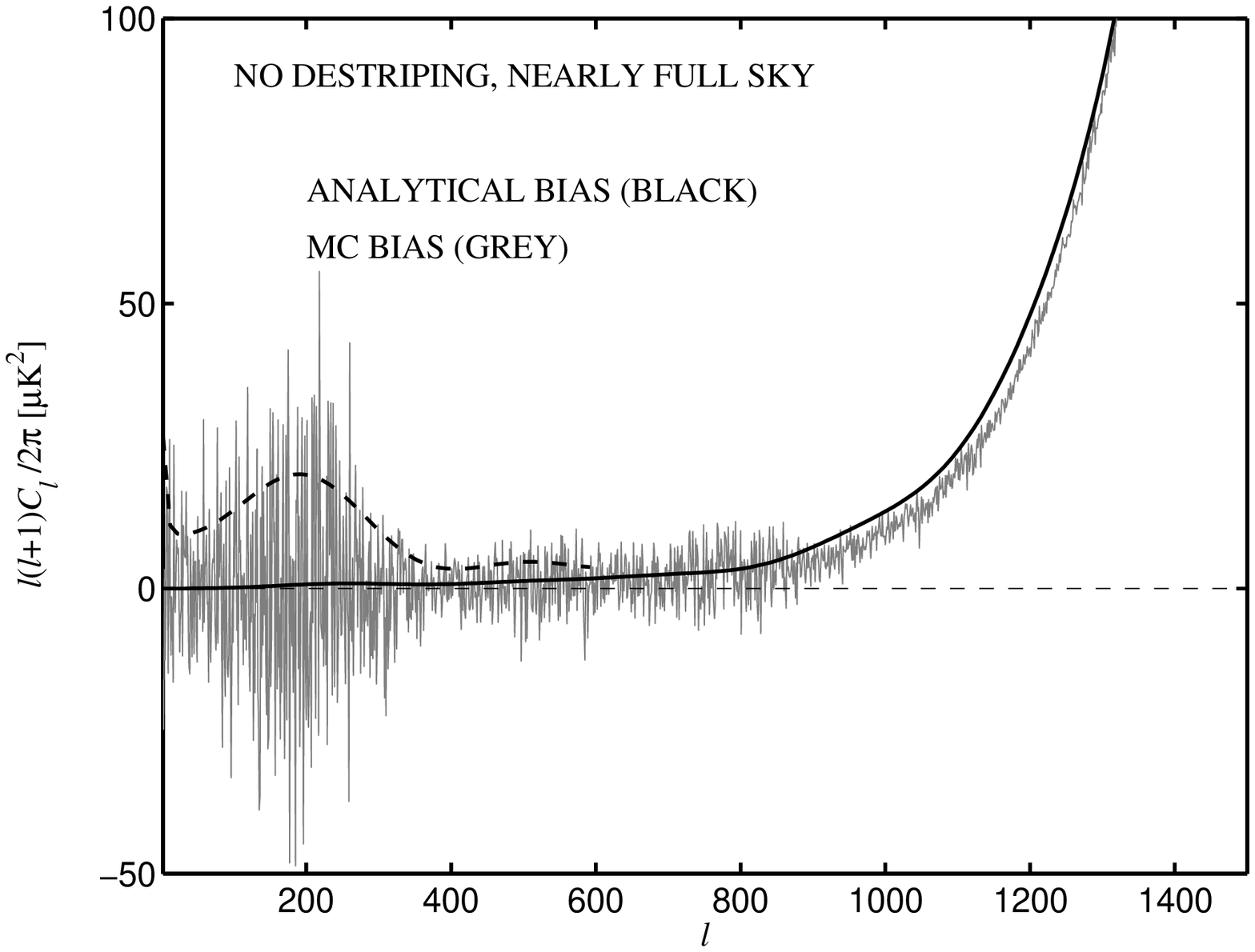}
\end{center}
\caption{{\bf Top panel}: Mean deconvolved pseudo power spectrum
$\sum_{\ell'}{M_{\ell \ell'}^{-1}\langle \widetilde{C}_{\ell'}
\rangle_{\mathrm{MC}}} / B^2_\ell$ (dashed curve), $\Lambda$CDM
input spectrum $C_\ell^{th}$ (solid upper curve) and deconvolved
signal bias estimate (difference of those two, solid lower curve).
Destriping has been applied in the map-making stage.
 {\bf Middle panel}: Deconvolved signal bias estimate shown with expanded
y-scale.
 {\bf Bottom panel}: Deconvolved signal bias estimate when no baseline removal
has been performed (no destriping). Also shown are the deconvolved
analytical approximation, Eq.\ (\ref{sl}), of the signal bias
$S_\ell^{app}$ (black curve) and the reference value $\Delta_{\rm
ref} \widehat{C}_\ell$ divided by $(N_{\rm MC}^{\rm (s)})^{1/2}$
(dashed curve), where $N_{\rm MC}^{\rm (s)} = 450$ is the number
of MC realizations to produce $\langle \widetilde{C}_\ell
\rangle_{\mathrm{MC}}$.
 }
 \label{s13}
\end{figure}

\begin{figure}
\begin{center}
\includegraphics[width=8cm,height=4.8cm]{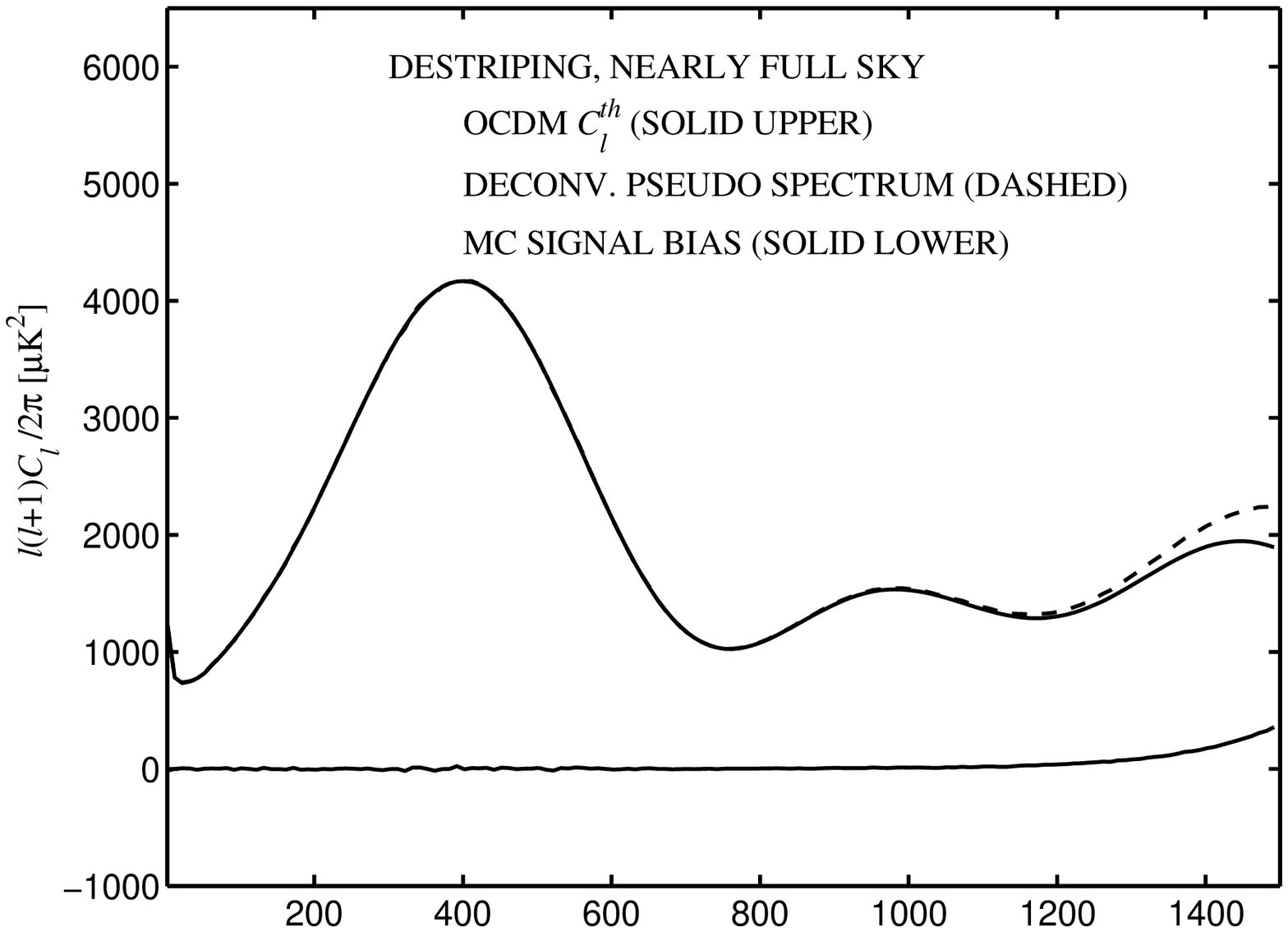}
\includegraphics[width=8cm,height=5cm]{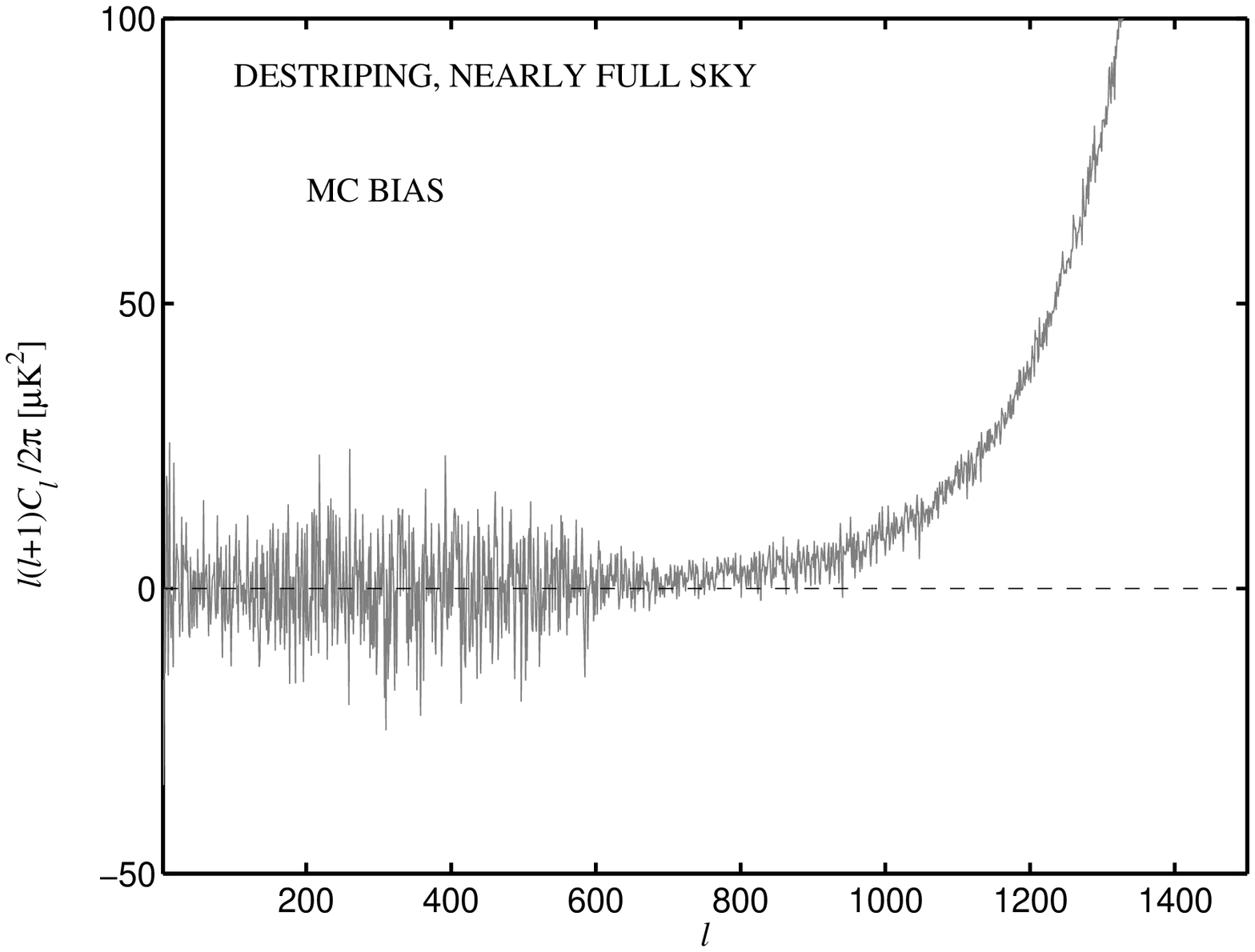}
\end{center}
\caption{Same as Fig.~\ref{s13} (two upper panels) but now the
cosmological model is OCDM.} \label{s13_open}
\end{figure}

We pick up the discussion from Section~\ref{sec:filter_function}, where we
introduced the signal bias in Eq.~(\ref{tildeT8}), reproduced here:
 \beq
   \langle \widetilde{C}_\ell \rangle = \sum_{\ell'} M_{\ell\ell'}
   B_{\ell'}^2 C_{\ell'}^{th} + S_\ell + \langle\widetilde{N}_\ell \rangle \,.
   \label{tildeT8again}
 \eeq
 The mean pseudo spectra ($\langle \widetilde{C_\ell}
\rangle_{\rm{MC}}$) obtained in Section~\ref{sec:filter_function}
were inserted in Eq.~(\ref{tildeT8again}) and the deconvolved
signal bias estimate $\sum_{\ell'}{M_{\ell
\ell'}^{-1}S_{\ell'}^{\mathrm{MC}} } / B^2_\ell$ was solved (with
$\langle \widetilde{N_\ell}\rangle = 0$). It is shown in
Fig.~\ref{s13} for the $\Lambda$CDM model and for the nearly full
sky. Fig.~\ref{s13_open} depicts similar things for the OCDM case.
The galactic cut cases look similar but somewhat noisier. The
signal bias estimates have a remarkable resemblance even though
the cosmological models are widely different. This indicates that
the signal bias approach is much better suited to our map-making
method than the filter function.

\subsection{Pointing Distribution}

Let us consider CMB only (no noise) and assume no baseline removal
for the time being. Let $k$ index the pixels of the final output
map. As a result of averaging the observations the output map
pixel temperature $\widetilde{T}_k$ of pixel $k$ is
 \beq
   \widetilde{T}_k = \frac{1}{p_k}\sum_{i=1}^{p_k}T_{ik}, \label{tildeT1}
 \eeq
where $p_k$ is the number of hits in pixel $k$, $i$ indexes a hit in pixel $k$
and $T_{ik}$ is the observed sky temperature of hit $i$. The pixel temperature
$\widetilde{T}_k$ can be written as
 \beq
   \widetilde{T}_k = T_k + \widetilde{T}_k - T_k = T_k + \Delta T_k\,,
   \label{tildeT2}
 \eeq
where $T_k$ is the beam and pixel smoothed sky temperature of pixel $k$.  The
temperatures $T_k$ and $T_{ik}$ can be given in terms of the expansion
coefficients $a_{\ell m}$ (see Eq. (\ref{harmofs}))
 \beq
   T_k = \sum_{\ell m} a_{\ell m} B_\ell Y_{\ell m}(\mathbf{n}_k)
   \label{tildeT3}
 \eeq
and
 \beq
   T_{ik} = \sum_{\ell m} a_{\ell m} B'_\ell Y_{\ell m}(\mathbf{n}_{ik}),
   \label{tildeT4}
 \eeq
where $B'_\ell$ contains the beam and pixel smoothing of the sky map and
$B_\ell$ contains the same for the output map. $\mathbf{n}_k$ is a unit vector
pointing to the center of the pixel $k$. $\mathbf{n}_{ik}$ is a unit vector of
the pointing of the observation $i$ in pixel $k$. Performing the reverse
spherical transform to both sides of Eq. (\ref{tildeT2}) one obtains a relation
for the expansion coefficients
 \begin{eqnarray}
   \Omega_p \sum_{k}\widetilde{T}_k Y_{\ell m}^\star(\mathbf{n}_k)
   = \Omega_p \sum_{k}T_k Y_{\ell m}^\star(\mathbf{n}_k)
   + \nonumber\\
   + \Omega_p \sum_{k} \Delta T_k Y_{\ell m}^\star(\mathbf{n}_k),
   \label{tildeT5}
 \end{eqnarray}
where only the hit pixels will be included in the sums.

The ensemble mean of the pseudo power spectrum
$\widetilde{C}_\ell$ can be obtained as
 \beq
   \langle \widetilde{C}_\ell \rangle = \frac{1}{2\ell+1} \sum_{m=-\ell}^{\ell}
   \langle |\Omega_p \sum_{k}\widetilde{T}_k Y_{\ell
   m}^\star(\mathbf{n}_k)|^2 \rangle. \label{tildeT6}
 \eeq
As a consequence of Eq. (\ref{tildeT5}) the mean can be split in
three terms
 \beq
   \langle \widetilde{C}_\ell \rangle
   = \langle C_\ell^x \rangle + \langle C_\ell^{\Delta} \rangle
   + \langle C_\ell^{x \Delta} \rangle \label{tildeT7}
 \eeq
with the following formulas
 \bea
   \langle C_\ell^x \rangle\!\! & = & \!\!\frac{\Omega_p^2}{2\ell+1}
   \sum_{m=-\ell}^{\ell} \sum_{k,p} \langle T_k T_p^\star \rangle Y_{\ell
   m}^\star(\mathbf{n}_k) Y_{\ell m}(\mathbf{n}_p) \,,
 \label{clx1} \\
   \langle C_\ell^{\Delta} \rangle\!\! & = & \!\!\frac{\Omega_p^2}{2\ell+1}
   \sum_{m=-\ell}^{\ell} \sum_{k,p} \langle \Delta T_k \Delta T_p^\star \rangle
   Y_{\ell m}^\star(\mathbf{n}_k) Y_{\ell m}(\mathbf{n}_p)
 \label{cld1}
 \eea
and
 \begin{eqnarray}
    \langle C_\ell^{x \Delta} \rangle = \frac{\Omega_p^2}{2\ell+1}
    \sum_{m=-\ell}^{\ell} \sum_{k,p} \big(\langle T_k \Delta T_p^\star
    \rangle Y_{\ell m}^\star(\mathbf{n}_k) Y_{\ell m}(\mathbf{n}_p) +{}
    \nonumber\\ {}+\langle T_k^\star \Delta T_p \rangle Y_{\ell m}(\mathbf{n}_k)
    Y_{\ell m}^\star(\mathbf{n}_p)\big).
   \label{clxd1}
 \end{eqnarray}
The double sum over pixel indices $k$ and $p$ extends only across
the hit pixels of the output map. Inserting Eq. (\ref{tildeT3})
and applying Eq. (\ref{clth}) one obtains for $\langle C_\ell^x
\rangle$
 \beq
   \langle C_\ell^x \rangle = \sum_{\ell'} M_{\ell\ell'}
   B_{\ell'}^2 C_{\ell'}^{th},
 \label{clx2}
 \eeq
where $M_{\ell\ell'}$ is the mode-mode coupling kernel
 \beq
   M_{\ell\ell'} = \frac{1}{2\ell+1}\sum_{m,m'=-\ell,-\ell'}^{\ell,\ell'}
   |\Omega_p \sum_{k} Y_{\ell m}^\star(\mathbf{n}_k)
   Y_{\ell' m'}(\mathbf{n}_k)|^2 \,.
 \label{mll}
 \eeq
The pseudo power spectrum $\langle \widetilde{C}_\ell \rangle$ can
thus be expressed as a sum of Eq. (\ref{clx2}) and two additional
terms
 \beq
   \langle \widetilde{C}_\ell \rangle = \sum_{\ell'} M_{\ell\ell'}
   B_{\ell'}^2 C_{\ell'}^{th} + \langle C_\ell^{\Delta} \rangle +
\langle C_\ell^{x \Delta} \rangle,
 \label{tildeT8b}
 \eeq
where we can identify $S_\ell = \langle C_\ell^{\Delta} \rangle +
\langle C_\ell^{x \Delta} \rangle$ as the signal bias defined in
Eq.\ (\ref{tildeT8}).

\subsection{Analytical Approximation to the Signal Bias}
\label{subsec:bias}

\subsubsection{Single Detector}
\label{subsubsec:single}

With the aid of some simplifying assumptions a reasonably accurate
analytical estimate for $S_\ell = \langle C_\ell^{\Delta} \rangle
+ \langle C_\ell^{x \Delta} \rangle$ (see Eqs. (\ref{cld1}) and
(\ref{clxd1})) can be derived. We consider here the case of a
single detector and nearly full sky coverage.

Let us consider an approximation to $\langle C_\ell^{\Delta}
\rangle$ first. This is called $S_\ell^{\Delta}$. There exists a
non-zero correlation between the temperatures $T_k$ of different
output map pixels. The temperature $\Delta T_k$ (see Eqs.
(\ref{tildeT1}) and (\ref{tildeT2})) is a linear combination of
differences between the temperatures of the hits falling into
pixel $k$ and the temperature at the centre point of that pixel.
The magnitudes of the pixel-to-pixel correlations are smaller for
these differences than for the temperatures themselves. Thus we
can apply the following approximation
 \beq
   \langle \Delta T_k \Delta T_p^\star \rangle = \langle |\Delta T_k|^2 \rangle
   \delta_{kp} \,.
   \label{sld1}
 \eeq
By applying the summation rules of the spherical harmonics we
obtain for $S_\ell^{\Delta}$
 \beq
   S_\ell^{\Delta} = \frac{\Omega_p}{n_{pix}} \sum_k \langle |\Delta T_k|^2 \rangle. \label{sld2}
 \eeq
The summation extends over the hit pixels of the output map. The
number of those pixels is $n_{pix}$. The formula exhibits no
$\ell$ dependence. The expectation value $\langle |\Delta T_k|^2
\rangle$ can be evaluated with the help of the input spectrum
$C_\ell^{th}$
 \beq
   \langle |\Delta T_k|^2 \rangle = \sum_{\ell=0}^{\ell_{max}}
   \frac{2\ell+1}{4\pi} C_\ell^{th} X_\ell^k,
 \label{sld5}
 \eeq
where
 \begin{eqnarray}
   X_\ell^k = \frac{B'_\ell B'_\ell}{p_k^2} \sum_{i,j=1}^{p_k} P_\ell(\cos(\theta_{ij}^k))
   - {} \nonumber\\ -{} \frac{2 B_\ell B'_\ell}{p_k}
   \sum_{i=1}^{p_k} P_\ell(\cos(\theta_{i}^k)) + B_\ell^2.
 \label{sld6}
 \end{eqnarray}
The upper limit for $\ell$ is $\ell_{max}$, $P_\ell(x)$ is the
Legendre polynomial, $\cos(\theta_{ij}^k) = \mathbf{n}_{ik} \cdot
\mathbf{n}_{jk}$ and $\cos(\theta_{i}^k) = \mathbf{n}_{ik} \cdot
\mathbf{n}_k$.

\begin{figure}
\begin{center}
\includegraphics[width=8cm,height=4.8cm]{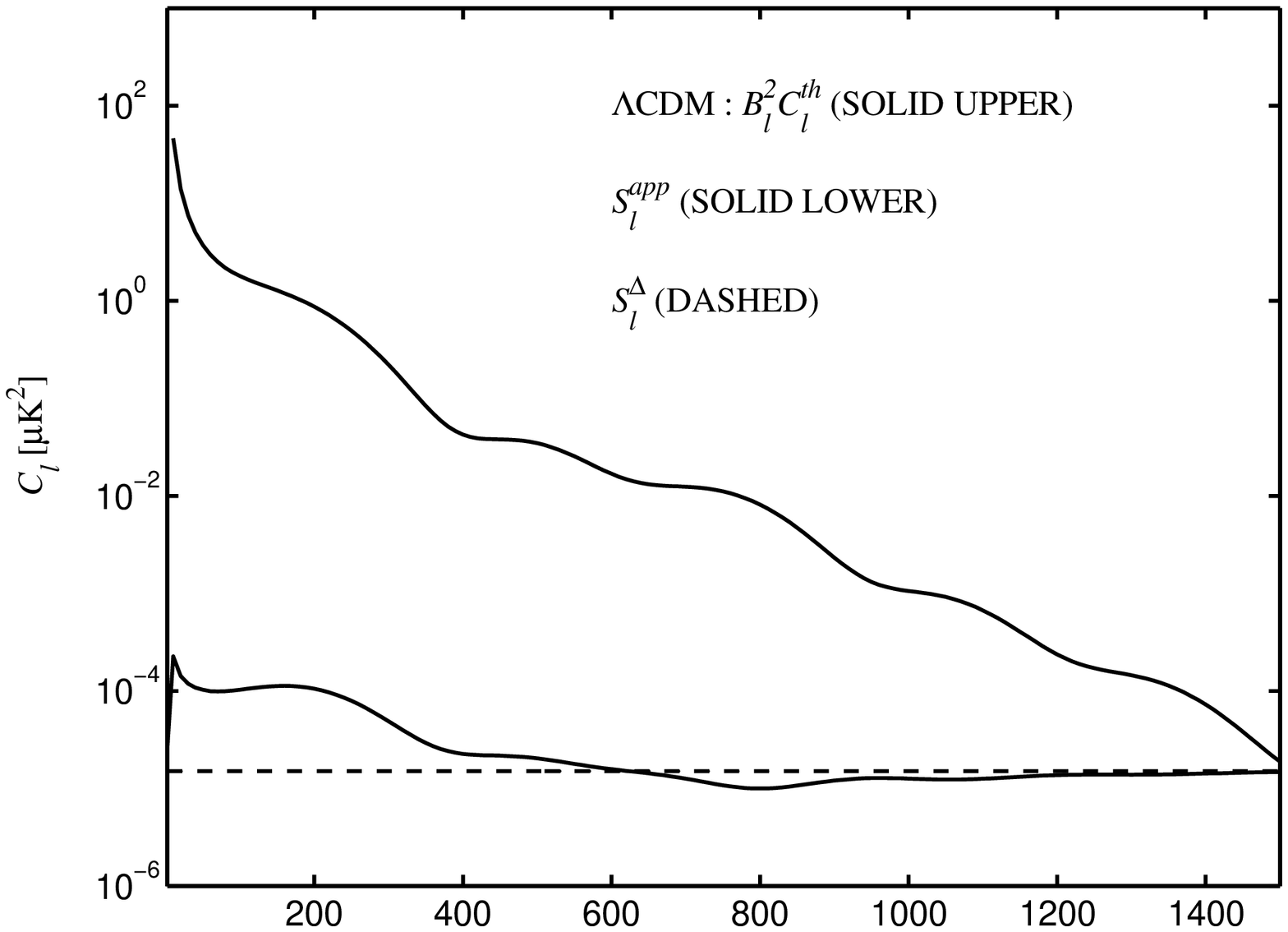}
\includegraphics[width=8cm,height=4.8cm]{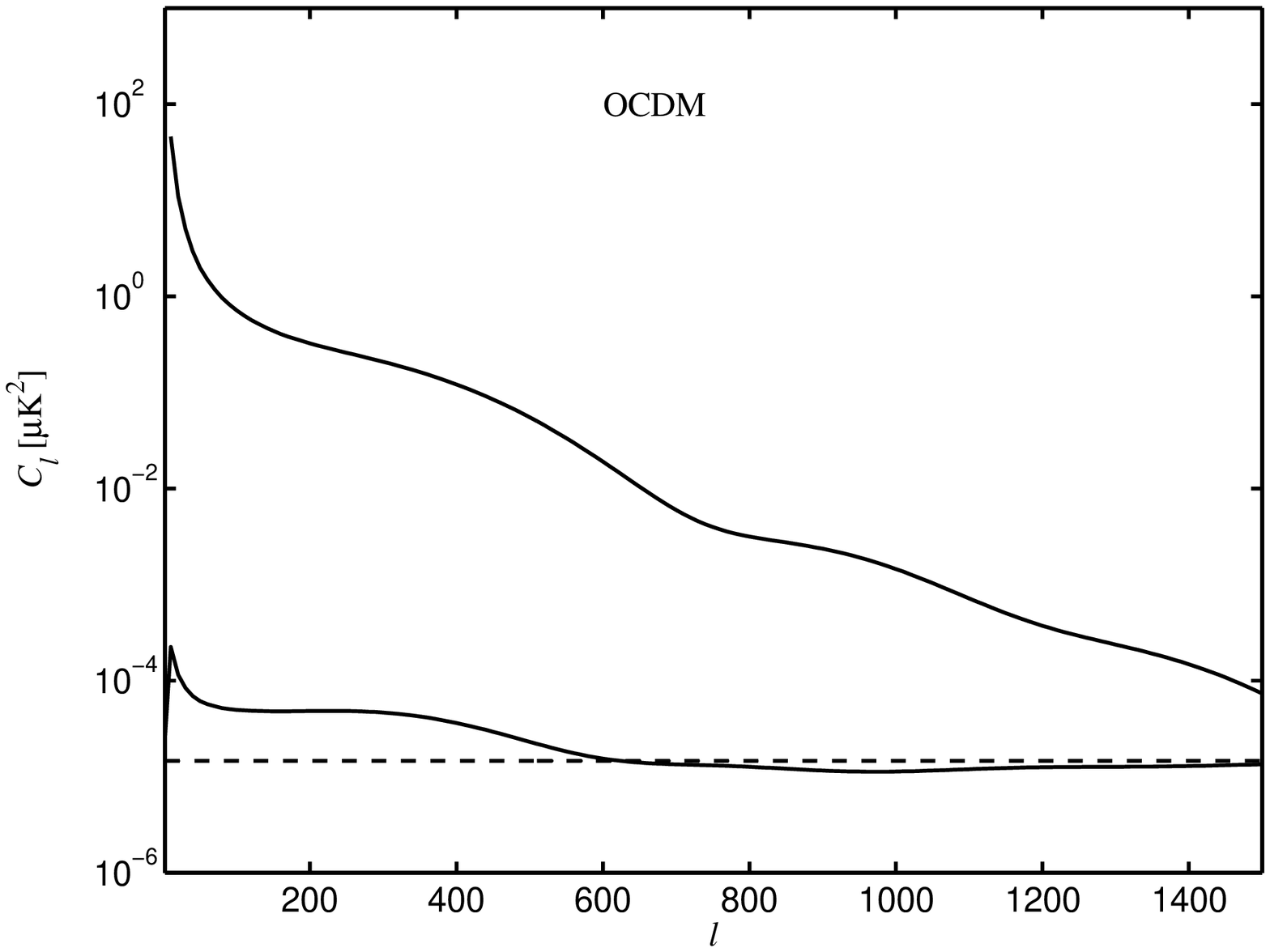}
\end{center}
\caption{Analytical approximations of the total signal bias
$S_\ell^{app}$ due to the detector pointing distribution (solid
lower curves) and the $\ell$-independent $S_\ell^{\Delta}$ part of
it (dashed curves).  {\bf Top panel} is for $\Lambda$CDM and {\bf
bottom panel} for OCDM.
The respective beam and pixel weighted CMB angular power spectra
are also shown (solid upper curves).} \label{slapp}
\end{figure}

By applying the pointings ($\mathbf{n}_{ik}$) of the 100 GHz LFI
detector scanning (see Section~\ref{sec:filter_function})
$S_\ell^{\Delta}$ was calculated and the result is shown in
Fig.~\ref{slapp} for $\Lambda$CDM and OCDM.

Let us examine the approximation of $\langle C_\ell^{x\Delta}
\rangle$ next. It is denoted by $S_\ell^{x\Delta}$. We shall
insert Eqs. (\ref{tildeT1}), (\ref{tildeT3}) and (\ref{tildeT4})
in Eq. (\ref{clxd1}) and use an approximation that is applicable
in the nearly full sky coverage
 \beq
   \Omega_p \sum_{k} Y_{\ell
   m}^\star(\mathbf{n}_k) Y_{\ell' m'}(\mathbf{n}_k) \simeq
   \delta_{\ell\ell'} \delta_{mm'}.
 \label{slxd1}
 \eeq
The summation is extended across the hit pixels. We obtain for
$S_\ell^{x\Delta}$
 \beq
     S_\ell^{x\Delta} = 2C_\ell^{th}\left(B_\ell B'_\ell U_\ell -
     B_\ell^2 \right),
 \label{slxd2}
 \eeq
where
 \beq
   U_\ell = \frac{1}{n_{pix}} \sum_{k} \frac{1}{p_k}
   \sum_{i=1}^{p_k} P_\ell(\cos(\theta_{i}^k)).
 \label{slxd3}
 \eeq

Finally we can write the analytical approximation
($S_\ell^\mathrm{app}$) of the signal bias $S_\ell$,
 \beq
  S_\ell^\mathrm{app} = S_\ell^{\Delta} + S_\ell^{x\Delta}.
  \label{sl}
 \eeq
It is shown in Fig.~\ref{slapp} for both $\Lambda$CDM and OCDM.
For low $\ell$ ($\ell < 400$) the signal bias is dominated by
$S_\ell^{x\Delta}$ and for high $\ell$ by $S_\ell^{\Delta}$. At
low multipoles the level of the signal bias is several orders of
magnitude lower than the level of the CMB power spectrum. At high
$\ell$ their levels become closer to each other.

As shown in the bottom panel of Fig.~\ref{s13} there is a good
match between the analytical approximation of the signal bias
$S_\ell^{app}$ and the signal bias estimate $S_\ell^{\mathrm{MC}}$
derived from the MC realizations.

\begin{figure}
\begin{center}
\includegraphics[width=8cm,height=4.8cm]{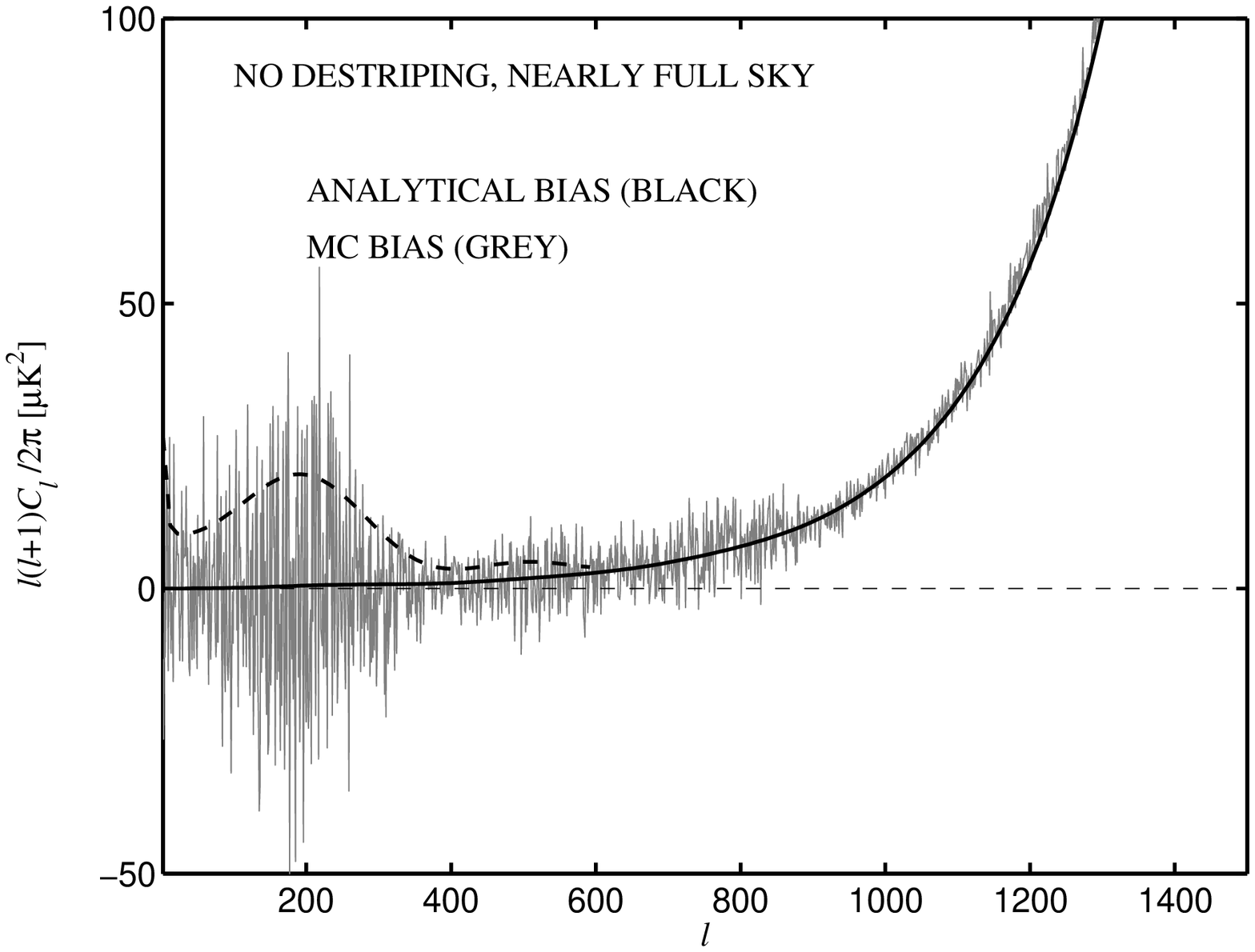}
\end{center}
\caption{Same as the bottom panel of Fig.~\ref{s13} but now the
input map and output map had higher resolutions ($N_\mathrm{side}
= 2048$ and $N_\mathrm{side} = 1024$, respectively).}
 \label{s1024}
\end{figure}

One might have expected that increasing the map resolution
(decreasing the pixel size) would have reduced the high-$\ell$
tail of the signal bias (and the filter function) or at least
would have shifted it to higher multipoles. Contrary to this
expectation the high-$\ell$ tail remains almost unaltered when the
map resolution is increased (see Fig.~\ref{s1024}).  We can now
explain this. The temperature differences $T_{ik} - T_k$ in
average and the pixel size $\Omega_p$ become smaller when the
pixel size is reduced. This would imply decreased $\Delta T_k$ and
$S_\ell^{\Delta}$. However, when the pixel size is reduced there
will be in average lower number of hits per pixel and those hits
will be more unevenly spread inside the pixel area. An uneven
spread would yield increased $\Delta T_k$. The net effect of these
two opposite phenomena is that the magnitude of $S_\ell^{\Delta}$
remains nearly unaltered in this case and the high-$\ell$ tail
remains roughly in the same place.

In other words, the crucial resolution element here is not the pixel size, but
the distance between neighboring detector pointings, separated by 3.3~arcmin
along the ring, the rings themselves being separated by 2.5~arcmin.

%

\subsubsection{Multiple Detectors}
\label{subsubsec:multi}

Destriping can be extended straightforwardly from a single detector TOD to
multiple TODs produced by several detectors. The baselines can be determined
jointly for all detectors and subtracted from the TODs. A single multidetector
map can be obtained by considering the hits of all detectors in a pixel. The
pseudo power spectrum can be extracted from the obtained multidetector map.

The signal bias will be altered when data from several detectors are combined,
since that increases the density of detector pointings. To study this issue the
pointing matrices of 70 GHz detectors of {\it PLANCK} LFI instrument were
available. There are totally six LFI detectors at 70 GHz. Each single detector
pointing matrix covered 12 months of observations. We calculated the analytical
approximation $S_\ell^{app}$ of the signal bias by utilizing Eqs. (\ref{sld2}),
(\ref{sld5}), (\ref{sld6}), (\ref{slxd2}) and (\ref{slxd3}). The number of
detectors was modelled by extending the calculations over the respective number
of pointing matrices. It should be noted that due to differences in the sky
coverage, in the mission time and in the sampling rates the absolute levels of
the signal biases of LFI 70 GHz and 100 GHz detectors cannot be directly
compared here. The results obtained with 70 GHz detectors reveal the relative
impact that the number of detectors will have on the signal bias of the pseudo
power spectrum extracted from a multidetector map.

\begin{figure}
\begin{center}
\includegraphics[width=8cm,height=4.8cm]{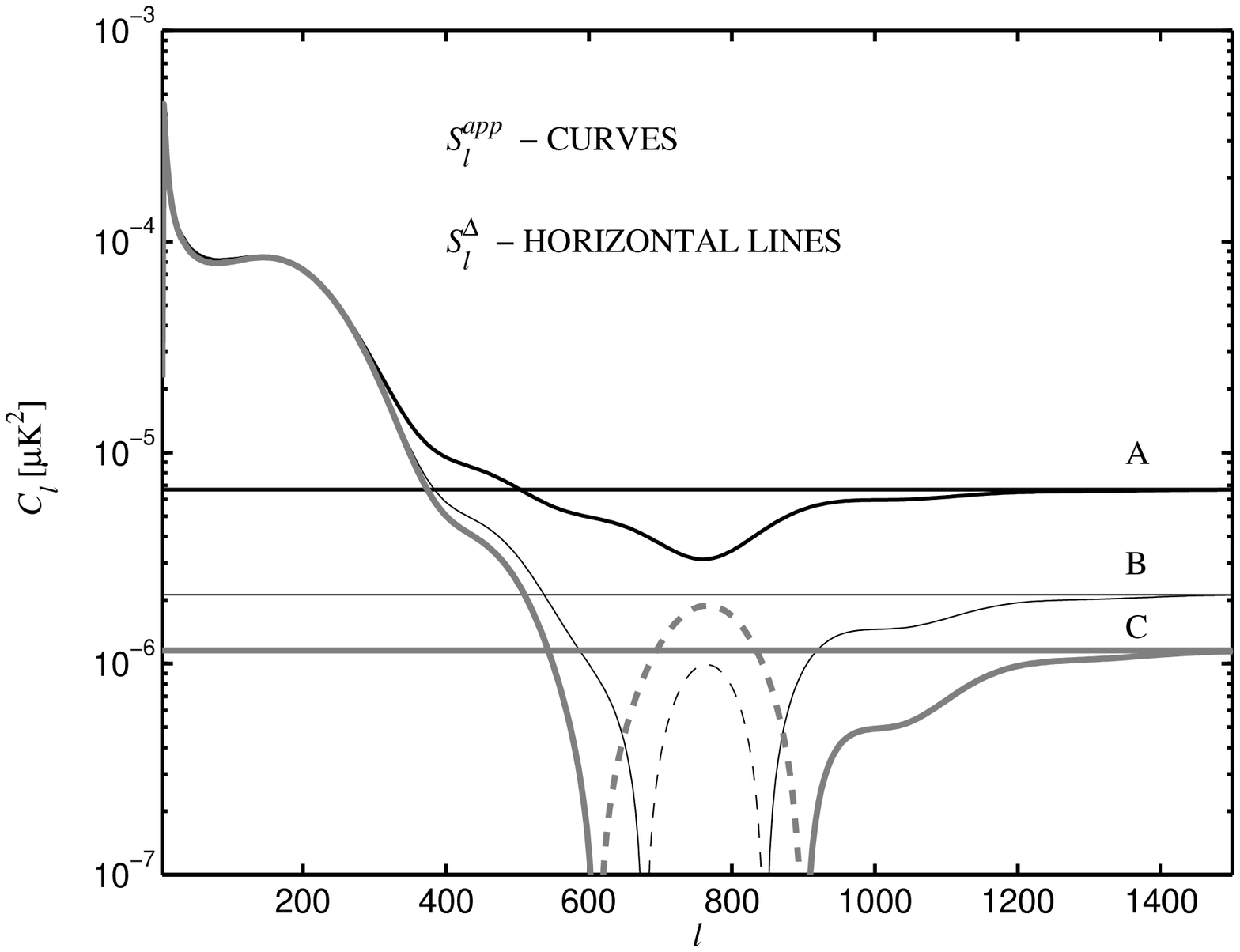}
\end{center}
\caption{Analytical approximations ($S_\ell^{app}$) of the signal
biases of one (A), three (B) and six (C) combined detector
signals. The bias of one detector is the mean of the signal biases
of the individual 70 GHz LFI detectors. The bias of three
detectors is obtained by considering the pointings of 70 GHz LFI
detectors 18, 19 and 20. The pixel size of the output map was
according to $N_\mathrm{side} = 512$ and all the detector
pointings fell in the centres of $N_\mathrm{side} = 1024$ pixels.
The horizontal lines show the contribution of
$S_\ell^{\Delta}$ in each case. Dashed lines indicate negative values of
$S_\ell^{app}$. The cosmological model was $\Lambda$CDM.}
 \label{sb_multdetector}
\end{figure}

The signal biases of one, three and six detectors are depicted in
Fig.~\ref{sb_multdetector}. At low $\ell$ ($\ell < 300$) the
signal bias is nearly independent of the number of detectors. The
absolute level of the signal bias is highest at this range of
multipoles. At high $\ell$ ($\ell > 1200$) the signal bias is
clearly dependent on the number of detectors. At this range of
multipoles the signal bias is dominated by $S_\ell^{\Delta}$. The
ratio between one detector and three detector $S_\ell^{\Delta}$ is
3.15. The respective ratio between one detector and six detector
$S_\ell^{\Delta}$ is 5.83. Thus the high-$\ell$ level of
$S_\ell^{app}$ is nearly inversely proportional to the number of
detectors involved. At the medium range of $\ell$'s the behavior
is more complex due to the abrupt change of sign of
$S_\ell^{x\Delta}$.

In light of this information, the level of the signal bias of a LFI 100 GHz
detector can be compared to the typical level of the angular power spectrum of
the instrument noise. The absolute level of the signal bias is highest at low
$\ell$ and this part is nearly independent of the number of detectors involved.
Comparing the analytical signal bias of Fig.~\ref{slapp} (single detector case)
to the angular spectrum of the instrument noise (shown for the modelled full
set of 24 detectors in the bottom panel of Fig.~\ref{clnoise}) one can notice
that in this case the low-$\ell$ level of the noise bias is at least couple of
orders of magnitude higher than the respective level of the signal bias.

\subsection{Estimating the Signal Bias}
\label{subsec:bias_or_filter}

In a real CMB experiment we will not have the true angular power
spectrum $C_\ell^{th}$ available for estimating $S_\ell$.  Since
we have concluded that its dependence on $C_\ell^{th}$ is weak,
this is not a concern. In place of $C_\ell^{th}$ we can use the
noisy power spectrum estimate $\widehat{C}_\ell$ that is obtained
from the real data.

In order to produce the noisy power spectrum estimate a single
realization of the signal+noise pseudo power spectrum
$\widetilde{C}_\ell$ was generated by following the procedure
described in Section~\ref{sec:times}. A single 100 GHz LFI
detector was considered. The estimate $\widehat{C}_\ell$ was
solved from Eq. (\ref{clestimate_sb}) by setting $S_\ell$ = 0 and
by using $\langle \widetilde{N}_\ell \rangle$ determined according
to the pure noise MC procedure (see Section~\ref{sec:times}). The
obtained $\widehat{C}_\ell$ was binned with $\Delta \ell$ = 10
bins at $\ell \le 1200$ and with $\Delta \ell$ = 50 bins at $\ell
> 1200$ and smoothed by a cubic spline (Fig.~\ref{sb_vs_ff_1},
top panel).

The noisy power spectrum estimate was used as the input spectrum
for signal-only MC simulations ($N_{\rm MC}^{(\rm s)} = 450$) to
produce the estimate of the signal bias (Fig.~\ref{sb_vs_ff_1},
bottom panel). Destriping was applied in the map-making. As
expected, the signal bias has a remarkable resemblance to the
signal bias obtained from simulations with the true $\Lambda$CDM
spectrum (see Fig.~\ref{s13}).


\label{lastpage}

\end{document}